\documentclass[mathpazo]{cicp}
\usepackage{mathrsfs}
\usepackage{subcaption}
\usepackage{amssymb}
\usepackage{amsbsy}
\usepackage[mathscr]{euscript}
\usepackage{amsmath}
\usepackage{mathrsfs}
\usepackage{subcaption}
\usepackage{amssymb}
\usepackage{amsbsy}
\usepackage{hyperref}
\usepackage{cleveref}
\usepackage{graphicx}
\usepackage[mathscr]{euscript}
\usepackage{comment}
\usepackage[T1]{fontenc}
\usepackage{color}

\numberwithin{equation}{section} 

\begin{document}


\title{Symmetric integration of the 1+1 Teukolsky equation on hyperboloidal foliations of Kerr spacetimes}      

\author[Markakis, C. et.~al.]{
Charalampos Markakis\affil{1,2}\comma\corrauth, 
Sean Bray\affil{1},
An{\i}l Zengino\u{g}lu\affil{3}
}
\address{
\affilnum{1}\ Mathematical Sciences, Queen Mary University of London, E1 4NS, London, UK\\
\affilnum{2}\ NCSA, University of Illinois at Urbana-Champaign, Urbana, Illinois 61801, USA\\
\affilnum{3}\ Institute for Physical Science and Technology, University of Maryland, College Park, Maryland 20742, USA\\
}
\email{{\tt c.markakis@qmul.ac.uk}}

\begin{abstract}
This work outlines a fast, high-precision  time-domain solver for scalar, electromagnetic and gravitational perturbations on hyperboloidal foliations of Kerr spacetimes.  Time-domain         Teukolsky equation solvers have typically used explicit         methods, which numerically violate Noether symmetries and are Courant-limited.         These restrictions can limit the performance of explicit schemes  when simulating  long-time extreme mass ratio inspirals,         expected to appear in LISA band for 2-5 years. We  thus explore         symmetric (exponential, Pad\'e   or Hermite) integrators, which are unconditionally stable and known to preserve certain Noether symmetries and phase-space volume.
For linear hyperbolic equations, these implicit integrators can be cast in explicit form, making them well-suited for long-time evolution of black hole perturbations.
The $1+1$ modal Teukolsky equation is discretized in space using         polynomial collocation methods and reduced to a linear system of ordinary differential equations, coupled  via  mode-coupling arrays and discretized (matrix) differential       operators. We use a matricization technique to cast the mode-coupled system in a form amenable to  a method-of-lines framework, which   simplifies   numerical implementation and enables  efficient parallelization on CPU and GPU architectures. 
 We test our numerical code by studying late-time tails of Kerr spacetime perturbations in the sub-extremal and extremal cases.
\end{abstract}

\ams{49S05, 49K20, 65M22, 65M25, 65M70, 70S10, 83-10, 83C25}
\keywords{Symmetric integration, Hermite integration, Pad\'e   methods, exponential integrator, Teukolsky equation, Aretakis instability, black hole perturbation theory, hyperboloidal slicing}

\maketitle

\section{Introduction}
\label{}
The computational cost of binary black hole inspiral simulations increases  proportionally to the square of the mass ratio $\mu=M/\mathfrak{m}\ge 1$, because one has to simulate for longer timescales with smaller time-steps as $\mu$ increases. Specifically, one factor of $\mu$ is due to the fact that the number of orbital cycles (and  the total time a signal is in the gravitational wave detector band)
is proportional to $\mu$. A  further factor of $\mu$ is due to the  
fact that, to resolve the disparate length scales  one needs to use a  finer spatial grid, which limits the maximum time-step allowed, due to the Courant-Friedrich-Lewy condition. This scaling makes traditional numerical relativity simulations computationally intractable for very large mass ratios $\mu$. Most accurate simulations are typically restricted to $\mu \lesssim 10$ (with a few recent simulations of a dozen orbital cycles prior to merger achieved for  binary systems with $\mu \sim 100$).

For intermediate ($10^{}\lesssim\mu\lesssim 10^{4}$) mass ratio inspirals, observable by LIGO, or extreme ($\mu\gtrsim 10^{4}$) mass ratio inspirals, observable by LISA, the use of  black hole perturbation theory becomes  favorable. The Einstein field equations are expanded in powers of a small quantity $q=1/\mu$, and the orbital dynamics are described by a point-particle, represented by a distributional (Dirac $\delta$-function)
source in the field equations. If the Dirac $\delta$-function is approximated, for instance, by a smooth Gaussian distribution, the fine spatial resolution requirement remains \cite{harmsNewGravitationalWave2014}, but discontinuous collocation methods \cite{oboyleConservativeEvolutionBlack2022}, discontinuous Galerkin methods \cite{fieldGPUAcceleratedMixedPrecisionWENO2021}, and effective source approaches \cite{vegaEffectiveSourceApproach2011,uptonSecondorderGravitationalSelfforce2021} reduce the disparity in scales and the need for a finer spatial grid, but a Courant limit is still present  if conditionally stable methods, such as explicit Runge-Kutta methods, are used. The presence of a Courant limit is especially restrictive for long-time simulations.

   In a suitable radiation gauge \cite{keidlGravitationalSelfforceRadiation2010,shahConservativeGravitationalSelfforce2011,shahFindingHighorderAnalytic2014,poundGravitationalSelfforceRadiationgauge2014,merlinCompletionMetricReconstruction2016}, the linearized Einstein field equations can take the form of the Teukolsky wave equation. Thus, in the context of black hole perturbation theory, gravitational self-force and LIGO and LISA waveform modelling,  one often numerically computes 
waveforms from  numerical solutions to the Teukolsky equation evaluated at null infinity. This  arises, for instance, when   evolving perturbations of a vacuum Kerr spacetime (simulating quasi-normal ringing and late-time tails after a binary black hole merger), or perturbations from a test-particle  on a geodesic orbit around a Kerr black hole, that is, simulating EMRIs to $0^{\rm{th}}$ order in the mass ratio. In these contexts, it has been shown that explicit Runge-Kutta methods fail to preserve energy, U(1) gauge charge, angular momentum, and phase-space volume when used to evolve black-hole perturbations in the time domain \cite{oboyleConservativeEvolutionBlack2022}. In an Extreme Mass Ratio Inspiral (EMRI), energy, angular momentum and unitarity will then be lost or gained artificially, to poor numerics, rather than lost purely to gravitational radiation. It has been further demonstrated \cite{oboyleConservativeEvolutionBlack2022} that the error in these quasi-conserved\footnote{For vacuum perturbations of compact support, these quantities are initially conserved on a spatial slice, until they reach the boundaries of the computational domain, whence they outflow through the black hole horizon or  null infinity.} quantities grows monotonically in time,  affecting the evolution not only quantitatively but also qualitatively. Using an explicit  Runge-Kutta method to solve a time-domain Teukolsky for  the perturbed Weyl scalar $\psi_4$ can thus result  in phase and amplitude error  in  waveforms, to  $0^{\rm{th}}$ order in the mass ratio. Moreover, if   these explicit Teukolsky solvers are used to compute  the Hertz potential $\Phi$ and the gravitational self-force in a self-consistent orbital evolution, the numerically simulated inspiral will be driven faster or slower than what gravitational-wave emission would account for. If the resulting self-forced particle orbit is used as a source to the Teukolsky equation for  $\psi_4$, the resulting waveform  can then suffer from phase and amplitude error accumulating  secularly to   $1^{\rm{st}}$ order in the mass ratio. 

In earlier work, it was also shown 
(cf. \cite{markakisDiscontinuousCollocationMethods2021,oboyleConservativeEvolutionBlack2022}) that time-symmetric integration methods are unconditionally stable and preserve energy and the U(1) gauge charge when used to evolve  linear perturbations of Schwarzschild black holes. Using such methods  to compute the gravitational self-force or radiation reaction from extreme mass ratio inspiral ensures that energy and angular momentum is lost only to true radiation, rather than numerical error, and yields the correct physical observables. This can be seen, for instance, by evolving a particle on a circular orbit and measuring the gravitational-wave phase and amplitude at  null infinity: symmetric methods lead to constant amplitude for constant orbital radius, while Runge-Kutta methods do not. Incorporating the perturbative radiation reaction fields, computed with time-symmetric methods,
in a self-consistent orbital evolution scheme will then lead to correct orbital characteristics, while Runge-Kutta methods lead to unphysically faster or slower inspiral. This further affects gravitational waveforms at higher perturbative orders. Given the important conservation and stability properties
of time-symmetric integration schemes, this work focuses on generalizing earlier results to the most astrophysically relevant case of Kerr spacetime \cite{kerrGravitationalFieldSpinning1963}. This work focuses on solving the homogeneous Teukolsky equation with symmetric methods, which serves as a  necessary step  towards incorporating distributional sources and simulating long-time EMRIs in the future.

This work uses geometric units $G=c=1$.

\section{1+1D Hyperboloidal Teukolsky Equation}
\label{method}
\subsection{Background}
Linear perturbations in Kerr spacetime can be described by  a set of  functions $\psi^{(s)}\in\mathbb{C}$ related to the spin-weight $s$ components of the Weyl tensor \cite{newmanApproachGravitationalRadiation1962}\footnote{That is, they transform as $\psi^{(s)} \rightarrow e^{i s \vartheta}~ \psi^{(s)}$ under frame rotations by an angle $\vartheta$ in the plane orthogonal to the radial direction.}.
Bardeen and Press have shown that such quantities obey a master wave equation in  Schwarzschild spacetime \cite{press_perturbations_1973,doi:10.1063/1.1666175}. This work was generalized to Kerr spacetime  by Press and Teukolsky \cite{teukolsky_perturbations_1973,teukolsky_perturbations_1974}. Bini et al. showed that the Teukolsky equation can be written in the 4D covariant form:
\begin{equation}
    g^{\mu \nu}(\nabla_\mu + s \Gamma_\mu)(\nabla_\nu + s \Gamma_\nu) \psi^{(s)} - 4 s^2 \Psi_2 \psi^{(s)} = 0
\end{equation}
where $\nabla_\mu$ is the covariant derivative compatible with the spacetime 4-metric $g_{\mu \nu}$, $\Gamma^\mu$ is a connection vector and $\Psi_2$ is the non-vanishing Weyl scalar for the unperturbed Type-D black hole spacetime (cf.~\cite{biniTeukolskyMasterEquation2002,toth_noether_2018} for explicit expressions). Setting $s=0$  recovers the Klein-Gordon equation for a complex scalar field and $s=\pm 1$  recovers  the Maxwell equations for electromagnetic test fields. Gravitational perturbations, corresponding to the linearized Einstein equations, are described by $s=\pm 2$. 
In Boyer-Lindquist coordinates  $(t,r,\theta,\phi)\in\{\mathbb{R} \times [M, \infty] \times \mathbb{S}^2\} $, the Teukolsky equation reads
\begin{equation} \label{TeukolskyMaster}
\begin{gathered}
  \left( {\frac{{{{({r^2} + {a^2})}^2}}}{\Delta } - {a^2}{{\sin }^2}\theta } \right)\partial _t^2{\psi ^{(s)}} + \frac{{4Mar}}{\Delta }{\partial _t}{\partial _\phi }{\psi ^{(s)}} + \left( {\frac{{{a^2}}}{\Delta } - \frac{1}{{{{\sin }^2}\theta }}} \right)\partial _\phi ^2{\psi ^{(s)}} \hfill \\
   - {\Delta ^{ - s}}{\partial _r}({\Delta ^{s + 1}}{\partial _r}{\psi ^{(s)}}) - \frac{1}{{\sin \theta }}{\partial _\theta }(\sin \theta {\partial _\theta }{\psi ^{(s)}}) + ({s^2}{\cot ^2}\theta  - s){\psi ^{(s)}} \hfill \\
   - 2s\left( {\frac{{M({r^2} - {a^2})}}{\Delta } - r - {\rm{i\,}}a\cos \theta } \right){\partial _t}{\psi ^{(s)}} - 2s\left( {\frac{{a(r - M)}}{\Delta } + \frac{{{\rm{i}}\cos \theta }}{{{{\sin }^2}\theta }}} \right){\partial _\phi }{\psi ^{(s)}} =0\hfill 
\end{gathered} 
\end{equation}
with ${\psi ^{(-s)}}$ satisfying the above equation with $s\rightarrow -s$.
Here,
  $a \in[-M,M]$ is the Kerr spin parameter,  
 $\Delta=(r-r_+)(r-r_-)$,
and 
\begin{equation} \label{horizons}
{r_ \pm } = M \pm {({M^2} - {a^2})^{1/2}}
\end{equation}
denotes the radial positions of the inner $(r_-)$ and outer $(r_+)$ horizon.

With numerical computation in mind, we will first perform a multipole expansion to separate the angular dependence into distinct modes and then express the Teukolsky equation as a coupled system of  1+1D partial differential equations in suitable time and radial coordinates.
To this end, we follow Barack et al. \cite{barackLateTimeDecay1999a,barackTimedomainMetricReconstruction2017} and expand the field $\psi^{(s)}$ in spin-weighted spherical harmonics ${}_sY_{\ell m} $ (cf. \cite{goldbergSpinSphericalHarmonics1967}):
\begin{equation} \label{multipolexpansion}
    \psi^{(s)}(t,r,\theta,\phi) =(r \Delta)^s \sum_{\ell=0}^\infty  \sum_{m = -\ell}^\ell \varphi_{\ell m}^{(s)} (t, r) {}_sY_{\ell m}(\theta,\phi_\star),
\end{equation}
where
\begin{equation}
 \phi_\star  = \phi+\int dr\frac{a}{\Delta}=\phi  + \frac{a}{{{r_ + } - {r_ - }}}\ln \left( {\frac{{r - {r_ + }}}{{r - {r_ - }}}} \right)
\end{equation}
is a horizon-regularized azimuthal coordinate  \cite{dolan_superradiant_2013,barackTimedomainMetricReconstruction2017}. 
The Kerr metric and the Teukolsky equation, when written in Boyer-Lindquist coordinates, have coordinate singularities at $r=r_\pm$. A tortoise radial coordinate $x$ can be defined by
\begin{equation} \label{Tortoisex}
x= \int dr \frac{{{r^2} + {a^2}}}{\Delta } =r - {r_ + } + \frac{{2M}}{{{r_ + }}}\ln \frac{{r - {r_ + }}}{{{r_ + } - {r_ - }}} - \frac{{2M}}{{{r_ - }}}\ln \frac{{r - {r_ - }}}{{{r_ + } - {r_ - }}}
\end{equation}
Inserting Eq.~\eqref{multipolexpansion} into the master equation \eqref{TeukolskyMaster} and using Clebsch-Gordan coefficients to re-expand the angular functions $ {}_sY_{\ell m} \cos \theta$ and $ {}_sY_{\ell m} \sin^2 \theta$ in spin-weighted spherical harmonics yields a 1+1D modal Teukolsky equation
\cite{barackLateTimeDecay1999a,dolan_superradiant_2013,barackTimedomainMetricReconstruction2017,longTimedomainMetricReconstruction2021}. 
In tortoise coordinates, this takes the form of a system of coupled 1+1 partial differential equations:
\begin{equation}\label{TeukolskyBarackModal}
\begin{gathered} 
  ({\partial _t^2}-{\partial_x^2})\varphi _{\ell m}^{(s)} + T(r){\partial _t}\varphi _{\ell m}^{(s)} + X(r){\partial _x}\varphi _{\ell m}^{(s)} + W(r)\varphi _{\ell m}^{(s)} \hfill \\
   - K(r)[{a^2}\partial _t^2(C_{ +  + }^{\ell }\varphi _{\ell  + 2,m}^{(s)} + C_ + ^{\ell }\varphi _{\ell  + 1,m}^{(s)} + C_0^{\ell m}\varphi _{\ell m}^{(s)} + C_ - ^{\ell }\varphi _{\ell  - 1,m}^{(s)} + C_{ -  - }^{\ell }\varphi _{\ell  - 2,m}^{(s)}) \hfill \\
   -8{\rm{\,i\,}}as{\partial _t}(c_ + ^{\ell }\varphi _{\ell  + 1,m}^{(s)} + c_0^{\ell }\varphi _{\ell ,m}^{(s)} + c_ - ^{\ell }\varphi _{\ell  - 1,m}^{(s)})] = 0 \hfill 
\end{gathered} 
\end{equation}
where $\partial_t$ and $\partial_x$ are taken with fixed $x$ and $t$ respectively. Due to  the axisymmetry of  Kerr spacetime, the $m$--modes are decoupled.  Unlike the Schwarzschild spacetime, however, due to the lack of spherical symmetry, the spherical harmonic $\ell$--modes in Kerr are coupled. Nevertheless, each mode  $\ell$ is coupled only to its nearest and next-to-nearest neighbours. The coupling coefficients are given by 
\begin{equation}
\begin{aligned}
  c_ - ^\ell & = \sqrt {\frac{{({\ell ^2} - {s^2})({\ell ^2} - {m^2})}}{{{\ell ^2}(4{\ell ^2} - 1)}}}   \\
  c_0^\ell  &=  - \frac{{ms}}{{\ell (\ell  + 1)}}  \\
  c_ + ^\ell & = c_ - ^{\ell  + 1} 
\end{aligned}
\end{equation}
and
\begin{equation}
\begin{aligned}
  C_{ \pm  \pm }^\ell  &=  -4 c_ \pm ^{\ell  \pm 1}c_ \pm ^\ell  \\ 
  C_ \pm ^\ell  &=  -4 c_ \pm ^\ell (c_0^{\ell  \pm 1} + c_0^\ell ) \\ 
  C_0^\ell  &=4[ 1 - {(c_ - ^\ell )^2} - {(c_ + ^\ell )^2} - {(c_0^\ell )^2}]
\end{aligned}
\end{equation}
where we  dropped the dependence of the coefficients on $m,s$ for brevity. Note that $c_ - ^\ell =0$ if $\ell=0$; $ c_0^\ell =0 $ if $\ell m s=0$;
$C_{--}^\ell=0$ if $|m| \le \ell \le |m|+1$ or  $|s| \le \ell \le |s|+1$;
$C_-^\ell=0$ if $\ell=0$ and
$C_0^\ell=0$ if $\ell<|s|$ or $\ell<|m|$. The radial functions in the above equation read
\begin{equation} \label{radialfunctions}
\begin{aligned}
K(r) &= \frac{\Delta }{{4{{({r^2} + {a^2})}^2}}}  \\ 
 T(r) &=16K(r)\{s(r + M) + {\rm{\,i\,}}{\mkern 1mu} asc_0^\ell  - 2Mr[s(r - M) - {\text{i}}{\mkern 1mu} am]/\Delta \}\  \\ 
 X(r) &= 16K(r)\{[s(r - M) - {\rm{\,i\,}}{\mkern 1mu} am)](1 + 2Mr/\Delta ) + {a^2}/r\}\  \\ 
W(r) &= 4K(r)[(\ell-s)(\ell+s+1)+2(s+1)M/r + 2{\rm{\,i\,}}am/2 -2 {a^2}/r^2 ]  
\end{aligned}
\end{equation}
These expressions may be obtained by transforming the 1+1D Teukolsky equation of Barack et al. \cite{barackLateTimeDecay1999a,barackTimedomainMetricReconstruction2017}
 from null to tortoise coordinates.

\subsection{Teukolsky equation in hyperboloidal coordinates}
\label{te_hyper}

For numerical computation purposes, it is beneficial to work with quantities of $\mathcal{O}(1)$, which can be accomplished by making the master equation dimensionless. Introducing rescaled tortoise  coordinates, \begin{equation} \label{trx4M}
t_\star= \frac{t}{4M}, \quad r_\star= \frac{x}{4M},
\end{equation}
and expressing the radial functions \eqref{radialfunctions} in terms of the compactified coordinate \eqref{sigma}, makes the modal equation \eqref{TeukolskyBarackModal}
dimensionless and, in particular, independent of the mass $M$.

We  use hyperboloidal slicing  \cite{Zenginoglu:2008wc,zenginogluIntermediateBehaviorKerr2014} to  handle the boundary conditions conveniently and accurately. The main computational advantages of  hyperboloidal slicing are that \textit{(i)} outflow boundary conditions are automatically satisfied on the future horizon  $\mathscr{H}^+$and future null infinity $\mathscr{I}^+$, \textit{(ii)} the finite computational domain  includes  $\mathscr{H}^+$  and  $\mathscr{I}^+$, and, consequently, \textit{(iii)} outgoing waveforms can be well-resolved throughout the domain and extracted at  $\mathscr{I}^+$ without extrapolation.  As shown in Ref.\cite{oboyleConservativeEvolutionBlack2022}, 
a transformation from tortoise coordinates $(t_\star,r_\star,\theta,\phi_\star)\in\{\mathbb{R}^2  \times \mathbb{S}^2\} $ to hyperboloidal coordinates $(\tau,\sigma,\theta,\phi_\star )\in\{\mathbb{R} \times [0, \sigma_+] \times \mathbb{S}^2\} $ can  be made using a height function $h$ and compactification function $g$ as follows:
\begin{subequations} \label{null_hyper}
\begin{eqnarray} \label{eq:null_hyper_1}
t_\star {\mkern-10mu} &=& {\mkern-10mu} \tau-h(\sigma), \\ \label{eq:null_hyper_2}
r_\star {\mkern-10mu} &=& \mkern-10mu g(\sigma).
\end{eqnarray}
\end{subequations}
Here, following Ref. \cite{schinkelInitialDataPerturbed2014,macedoHyperboloidalFrameworkKerr2020,oboyleConservativeEvolutionBlack2022,ripleyComputingQuasinormalModes2022}, we introduce a  radial coordinate
\begin{equation} \label{sigma}
\sigma= \frac{2M}{r}
\end{equation}
in a compactified radial domain $\sigma \in [0,\sigma_+]$, with future null infinity $\mathscr{I}^+$ located at   $\sigma_{\mathscr{I}^+}=0$ and the future horizon $\mathscr{H}^+$ (cf.~Eq.~\eqref{horizons}) located at 
\begin{equation}
 {\sigma _{{\mathcal{H}^ + }}} =\sigma_+ := \frac{{2M}}{{{r_ + }}} = \frac{2}{{1 + \sqrt {1 - 4{\chi ^2}} }}
\end{equation}
Then, inserting Eqs.~\eqref{sigma} and \eqref{trx4M} into Eq.~\eqref{Tortoisex}
  determines the compactification function \eqref{eq:null_hyper_2}:
\begin{eqnarray}\label{eq:height_f}
g(\sigma) \mkern-10mu &=& \mkern-10mu \int \frac{{1 + {\sigma ^2}{\chi ^2}}}{{2{\sigma ^2}( \sigma (1 - \sigma {\chi ^2} )-1)}}
\nonumber\\ \mkern-10mu
&=& \mkern-10mu \frac{1}{2} \left[ {\frac{1}{\sigma } + \frac{1}{{1 - {\kappa ^2}}}\log \left( {\frac{{1 + {\kappa ^2}}}{\sigma } - 1} \right) - \frac{{{\kappa ^2}}}{{1 - {\kappa ^2}}}\log \left( {\frac{{1 + {\kappa ^2}}}{\sigma } - {\kappa ^2}} \right)} \right]  \quad  \quad
\end{eqnarray}
where we defined dimensionless spin parameters $\kappa\in[-1,1]$ and $\chi\in\big[-\frac{1}{2},\frac{1}{2}\big]$  given by 
\begin{eqnarray}\label{eq:kappa}
\kappa \mkern-10mu &:=& \mkern-10mu\frac{a/M}{{1 + \sqrt {1 - {(a/M)^2}} }}\\
\chi\mkern-10mu  &:=& \mkern-10mu\frac{a}{2M} \label{eq:chi}
\end{eqnarray}
Eq.~\eqref{eq:height_f} may appear to be more complicated than other common choices of compactification function. 
However,  most choices typically lead to a final form of the Teukolsky equation  in hyperboloidal slicing  which is several pages long  \cite{zenginogluSpacelikeMatchingNull2009,raczNumericalInvestigationLatetime2011,harmsNumericalSolutionTeukolsky2013,csukasNumericalInvestigationDynamics2019,fieldGPUAcceleratedMixedPrecisionWENO2021}. 
Because the compactification function~\eqref{eq:height_f} has a   simple derivative, and follows  from the  very simple compactification \eqref{sigma} in conjunction with
the tortoise coordinate~\eqref{Tortoisex} (and the latter is known to simplify  Kerr perturbation equations),   it is expected to simplify the final form of the Teukolsky equation.
The height function in Eq.~\eqref{eq:null_hyper_1} can be determined
by integrating outgoing null geodesics \cite{Schinkel:2013tka,schinkelInitialDataPerturbed2014,ripleyComputingQuasinormalModes2022}
asymptotically  and requiring that level sets $\Sigma_\tau$ of the time coordinate $\tau$ become null hypersurfaces near future null infinity  $r\rightarrow \infty \Leftrightarrow\sigma \rightarrow 0$. Keeping the first three terms in this asymptotic expansion yields
\begin{equation}\label{eq:compact_f}
h(\sigma) = g(\sigma) - \frac{1}{\sigma } + \log \sigma + \sigma +\mathcal{O}(\sigma^2).  \end{equation}
Dropping terms of $\mathcal{O}(\sigma) $  and higher amounts to the so-called \textit{minimal gauge} \cite{macedo_axisymmetric_2014,macedoHyperboloidalFrameworkKerr2020}. 
This gauge retains the minimal structure in the coordinate transformation needed to construct hyperboloidal slices. Consequently, the corresponding equations of black-hole perturbation theory assume their  simplest form. 
Another way to obtain the height function is as follows.
Given the tortoise coordinates~\eqref{trx4M}, the radially in- and outgoing characteristics are  $t_\star \pm r_\star$. With the compactification~\eqref{eq:height_f}, the leading singular behavior near future null infinity  is
\begin{equation} \label{eq:limit_g} \nonumber
g_{\mathscr{I}^+}(\sigma )\approx\frac{1}{2} \left(  \frac{1}{\sigma } - \log \sigma -\sigma  \right).  \end{equation}   
Terms of $\mathcal{O}(\sigma) $ and higher are regular at $\sigma=0$ and may be omitted. We seek a time coordinate $\tau = t + h(\sigma)$ that is ingoing near the future horizon, $h_{\mathscr{H}^+}(\sigma) \approx r_\star = g(\sigma)$, and outgoing near future null infinity, $h_{\mathscr{I}^+}(\sigma) \approx -r_\star = -g(\sigma)$. There is an infinite dimensional space of choices that satisfy these asymptotic conditions.  The \textit{minimal gauge} sets the ingoing behavior as $\tau \approx t+g(\sigma)$. Then it corrects the behavior at infinity by subtracting twice the leading infinity terms:
\begin{equation}\label{eq:h_Anil} \nonumber
h(\sigma) = g(\sigma) -2 g_{\mathscr{I}^+}(\sigma)=g(\sigma)-\frac{1}{\sigma}+\log \sigma +\sigma + \mathcal{O}(\sigma^2).  \end{equation}
Then, to leading singular order near infinity, we obtain the desired behavior
\begin{equation}\label{eq:hinfy_Anil} \nonumber
h_{\mathscr{I}^+}(\sigma) \approx  - g_{\mathscr{I}^+}(\sigma).
\end{equation}
\begin{figure}[ht]
        \centering
        \includegraphics[scale=1]{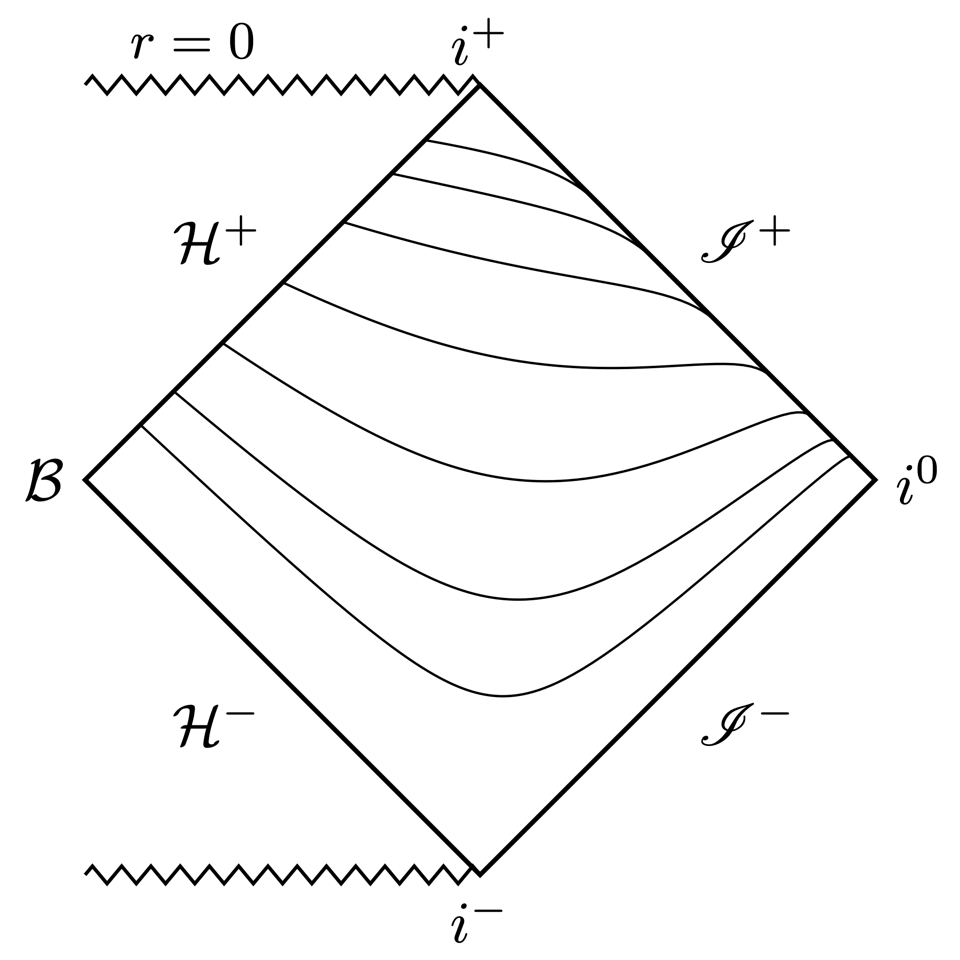}
        \caption{Carter-Penrose diagram for the black hole exterior in the Schwarzschild limit. The solid black curves depict hyperboloidal $\tau=$constant surfaces extending between the future horizon $\mathscr{H}^+$ at $\sigma=\sigma_+$ and future null infinity $\mathscr{I}^+$ at $\sigma =0$. The hyperboloidal coordinates $(\tau,\sigma)\in\{\mathbb{R} \times [0, \sigma_+]  \} $
 smoothly foliate the entire black hole exterior so that both $\mathscr{H}^+$ and $\mathscr{I}^+$ are part of the finite computational domain. As light cones tip over at these boundaries, outflow boundary conditions are naturally satisfied automatically in these coordinates.} 
        \label{fig:penrose_diagram}
\end{figure}
When we perform the coordinate transformation \eqref{null_hyper} on Eq.~\eqref{TeukolskyBarackModal} and impose minimal gauge, we  obtain a remarkably simple hyperboloidal 1+1D Teukolsky equation:
\begin{equation} \label{Teukolsky11hyper}
\begin{gathered}
  (1 + \sigma )(1 - {\chi ^2}\sigma )\partial _\tau ^2\varphi _\ell^{(s)} + (s(\sigma  - 1) - {\chi ^2}\sigma (1 + 3\sigma ) + 2\sigma  + {\text{i}}m\chi (1 + 2\sigma )){\partial _\tau }\varphi _\ell ^{(s)}   \hfill \\
  - (1 - {\sigma ^2}(2 - {\chi ^2}(1 + 2\sigma ))){\partial _\tau }{\partial _\sigma }\varphi _\ell^{(s)} + \sigma (s(\sigma  - 2) - 2 + \sigma (3 + 2{\text{i}}m\chi  - 4{\chi ^2}\sigma )){\partial _\sigma }\varphi_\ell^{(s)} \hfill \\
   - {\sigma ^2}(1 - \sigma  + {\chi ^2}{\sigma ^2})\partial _\sigma ^2\varphi _\ell^{(s)} + [
\ell(\ell - s)(\ell + 1 + s)
 + (1 + s + 2{\text{i}}m\chi )\sigma  - 2{\chi ^2}{\sigma ^2}]\varphi_\ell^{(s)} \hfill \\ 
  - {\chi ^2}\partial _\tau ^2({\text{C}}_{ -  - }^\ell\varphi_{\ell - 2}^{(s)} + {\text{C}}_ - ^\ell\varphi _{\ell - 1}^{(s)} + {\text{C}}_0^\ell\varphi _l^{(s)} + {\text{C}}_ + ^\ell\varphi_{\ell + 1}^{(s)} + {\text{C}}_{ +  + }^l\varphi_{l + 2}^{(s)}) \hfill\\ +{\text{i}}s\chi {\partial _\tau }(c_ - ^\ell\varphi_{\ell - 1}^{(s)} + c_0^\varphi{_\ell}^{(s)} + c_ + ^\ell\varphi _{\ell + 1}^{(s)}) = 0 \hfill \\ 
\end{gathered} 
\end{equation}
where  the dependence of the field $\varphi _{\ell m}^{(s)}$ on the index $m$ has been suppressed for brevity. 
Eq.~\eqref{Teukolsky11hyper} is polynomial in the radial coordinate $\sigma$ and, to our knowledge, is the simplest form of the hyperboloidal  Teukolsky equation. 
Eq.~\eqref{Teukolsky11hyper} generalizes the Bardeen-Press-Teukolsky equation in minimal gauge to Kerr  spacetime. The Schwarzschild limit $\chi \rightarrow0$ of the above equation is in agreement with     
Ref.\cite{jaramilloPseudospectrumBlackHole2021,oboyleConservativeEvolutionBlack2022}. Eq.~\eqref{Teukolsky11hyper} satisfies outflow boundary conditions at   $\mathscr{H}^+$  and  $\mathscr{I}^+$. For instance, if we  set $\sigma=0$ in Eq.~\eqref{Teukolsky11hyper}  and consider the terms with the highest derivatives, we obtain an operator with the  structure 
$(1-\chi^2 C_0^\ell) \partial_\tau^2 - \partial_\tau \partial_\sigma$ 
for the $s,\ell$ term. In particular, all $\partial_\sigma$ and $\partial_\sigma^2$ terms vanish at $\sigma=0$. This implies  that there are no incoming boundary conditions, and the outgoing characteristic speed is positive. We will return to the topic of boundary conditions after computing the principal symbol below.

To address  the $\ell$--mode coupling, let us write Eq.~\eqref{Teukolsky11hyper} in matrix form as follows
\begin{equation}\label{eq:hyper_barak}
 \sum_{k=|s|}^{\ell_{\max}}        \left( {{{\rm A}_{\ell k}}\partial _\tau ^2 - {{\rm B}_{\ell k}}{\partial _\tau } - {\Gamma _{ \ell k}}{\partial _\sigma }{\partial _\tau } - {\Delta _{\ell k}}{\partial _\sigma } - {{\rm E}_{\ell k}}\partial _\sigma ^2 - {{\rm Z}_{\ell k}}} \right){\varphi _k^{(s)} } = 0 
\end{equation}
where ${\ell_{\max}}>|s|$ is the maximum $\ell$--multipole one wishes to keep in the spherical harmonic expansion.
The terms ${{\rm A}_{\ell k}}, {\rm B}_{\ell k},  {\Gamma _{\ell k}},\Delta _{\ell k}
,{\rm E}_{\ell k},{\rm Z}_{\ell k}$ are given by
\begin{equation}\label{eq:2}
\begin{gathered}
{{\rm A}_{\ell k}} (\sigma) = \left( {(1 + \sigma)(1  - {\chi^2}\sigma )- {\chi ^2}{C_0^{\ell}}} \right) \delta _{k \ell} \hfill \\
\hfill- {\chi ^2}({\delta _{k,\ell - 2}}{C_{--}^{\ell}}+ {}{\delta _{k,\ell - 1}}{C_-^{\ell}}  + {\delta _{k,\ell+1}}{C_+^{\ell} } +{\delta _{k,\ell+2}}{C_{ +  + }^{\ell}})  \\ 
{\rm B} _{\ell k} (\sigma)= \left(s(1- \sigma) -2 \sigma+ \chi ^2\sigma 
   (1+3 \sigma )-{\rm{i}} m \chi (1+2 \sigma)-{\rm{i}} s \chi c_0^\ell  \right)  \delta _{k \ell} \\
   \hfill -{\rm{i}} s \chi(  \delta _{k,\ell-1} c_-^\ell+  \delta _{k,\ell+1} c_+^\ell)
\\
\Gamma _{\ell k} (\sigma)=  \left( {1 - {\sigma ^2}\left( {  2 - \left( {1 + 2\sigma } \right){\chi ^2}} \right)} \right){\delta _{k \ell}} \hfill \\ 
\Delta _{\ell k}
                (\sigma) =   \sigma \left( {  2 + s\left( {  2- \sigma } \right) - \sigma \left( {3   - 4 {\chi ^2} \sigma+ 2{\rm{i}}m\chi} \right)} \right){\delta _{k \ell}}  \hfill   \\
                {\rm E}_{\ell k}(\sigma)= {\sigma ^2}\left( {1 - \sigma \left( {  1- {\chi ^2}\sigma } \right)} \right){\delta _{k \ell}} \hfill \\
{\rm Z}_{\ell k}  (\sigma)= \left( (s- \ell )(\ell + 1 + s) - (1 + s + 2{\text{i}}m\chi )\sigma  + 2{\chi ^2}{\sigma ^2} \right){\delta _{k \ell}} \hfill
\end{gathered}
\end{equation}
where $\delta _{k \ell}$ denotes the Kronecker delta. 

By introducing the auxiliary variable $\pi_\ell^{(s)}:= \partial_\tau \varphi_\ell^{(s)} $\footnote{Some numerical methods may benefit from other  auxiliary variable choices, such as canonical variables. Certain choices   can  reduce the system to first-order in space without introducing  a second auxiliary variable. We will limit ourselves to the simplest choice here.}, Eq.~\eqref{eq:hyper_barak} takes the first-order in time,~second-order in space form:
\begin{equation} \label{time1space2}
 {\Bigg( \mkern-8mu {\begin{array}{*{20}{c}}
  {{\delta _{\ell k}}}&0 \\ 
  0&{{{\rm A}_{\ell k}}} 
\end{array}} \mkern-8mu \Bigg){\partial_\tau }\Bigg( \mkern-8mu{\begin{array}{*{20}{c}}
  {\varphi _k^{(s)}} \\ 
  {\pi _k^{(s)}} 
\end{array}}  \mkern-8mu \Bigg)}  = {\Bigg( \mkern-8mu {\begin{array}{*{20}{c}}
  0&{{\delta _{\ell k}}} \\ 
  {{\Delta _{\ell k}}{\partial _\sigma } + {{\rm E}_{\ell k}}\partial _\sigma ^2 + {{\rm Z}_{\ell k}}}&{{{\rm B}_{\ell k}} + {\Gamma _{\ell k}}{\partial _\sigma }} 
\end{array}} \mkern-8mu \Bigg)\Bigg( \mkern-8mu {\begin{array}{*{20}{c}}
  {\varphi _k^{(s)}} \\ 
  {\pi _k^{(s)}} 
\end{array}}  \mkern-8mu \Bigg)} \qquad
\end{equation}
where summation over $k=|s|, \dots, \ell_{\max}$ is implied on both sides of this  equation. 

 A first-order in space reduction may be obtained by defining 
a second auxiliary variable  $p_\ell^{(s)}:= \partial_\sigma \varphi_\ell^{(s)} $, which brings Eq.~\eqref{eq:hyper_barak} to the fully first-order form:
\begin{equation}   \label{time1space1}
\begin{gathered}
  \left( {\begin{array}{*{20}{c}}
  {{\delta _{\ell k}}}&0&0 \\ 
  0&{{{\text{A}}_{\ell k}}}&0 \\ 
  0&0&{{\delta _{\ell k}}} 
\end{array}} \right){\partial _\tau }\left( {\begin{array}{*{20}{c}}
  {\varphi _k^{(s)}} \\ 
  {\pi _k^{(s)}} \\ 
  {p_k^{(s)}} 
\end{array}} \right) + \left( {\begin{array}{*{20}{c}}
  0&0&0 \\ 
  0&{ - {\Gamma _{\ell k}}}&{{{\text{E}}_{\ell k}}} \\ 
  0&{{\delta _{\ell k}}}&0 
\end{array}} \right){\partial _\sigma }\left( {\begin{array}{*{20}{c}}
  {\varphi _k^{(s)}} \\ 
  {\pi _k^{(s)}} \\ 
  {p_k^{(s)}} 
\end{array}} \right) \\ 
  \hfill  + \left( {\begin{array}{*{20}{c}}
  0&{{\delta _{\ell k}}}&0 \\ 
  {{{\text{Z}}_{\ell k}}}&{ - {{\text{B}}_{\ell k}}}&{{\Delta _{\ell k}}} \\ 
  0&0&0 
\end{array}} \right)\left( {\begin{array}{*{20}{c}}
  {\varphi _k^{(s)}} \\ 
  {\pi _k^{(s)}} \\ 
  {p_k^{(s)}} 
\end{array}} \right) = 0 \\ 
\end{gathered} 
\end{equation} 
A disadvantage of the system \eqref{time1space1} is that the constraint
  $p_\ell^{(s)}- \partial_\sigma \varphi_\ell^{(s)}=0 $ is numerically violated during time evolution, so constraint damping is typically required \cite{deppeCriticalBehavior3d2019}.
For this reason,  our numerical scheme will be based on the second-order in space system \eqref{time1space2}. 
The system \eqref{time1space1} is nevertheless useful for computing the principal symbol and the characteristic speeds in order to study  hyperbolicity and  boundary conditions. The ingoing speeds $\lambda_{\rm in}^{|s|},\lambda_{\rm in}^{{|s|}+1},\dots,\lambda_{\rm in}^{\ell_{\max}}$ and the outgoing speeds $\lambda_{\rm out}^{|s|},\lambda_{\rm out}^{|s|+1},\dots,\lambda_{\rm out}^{\ell_{\max}}$
are the eigenvalues of the $2(\ell_{\max}-|s|) \times  2(\ell_{\max}-|s|)  $ matrix
\begin{equation}
{\left( {\begin{array}{*{20}{c}}
  { - {\text{A}}_{\ell k}^{ - 1}{\Gamma _{\ell k}}}&{{\text{A}}_{\ell k}^{ - 1}{{\text{E}}_{\ell k}}} \\ 
  {{\delta _{\ell k}}}&0 
\end{array}} \right)},  \quad   k,\ell=|s|,\dots,\ell_{\max}
\end{equation}
and satisfy the characteristic determinant equation\footnote{Multiplying the characteristic equation with the matrix inverse ${{{\text{A}}_{\ell k}^{-1}}}$ does not change the eigenvalues, cf. e.g. \cite{papadopoulosRelativisticHydrodynamicsSpacelike1999}}:
\begin{equation}
\left| {\left( {\begin{array}{*{20}{c}}
  { - {\Gamma _{\ell k}}}&{{{\text{E}}_{\ell k}}} \\ 
  {{\delta _{\ell k}}}&0 
\end{array}} \right) - \lambda \left( {\begin{array}{*{20}{c}}
  {{{\text{A}}_{\ell k}}}&0 \\ 
  0&{{\delta _{\ell k}}} 
\end{array}} \right)} \right| = 0.
\end{equation}
Due to the $\ell$--mode coupling, the matrix ${\text{A}}_{\ell k}(\sigma)$ is not diagonal and a simple closed form expression for the speeds $\lambda(\sigma)$ is not available (except for a small number of modes $\ell_{\max}$
or in the Schwarzschild limit $\chi = 0$). 
Nevertheless, it is possible to obtain series approximations to $\lambda(\sigma)$ near the horizon
  $\sigma ={\sigma _{{ + }}} $
 or future null infinity 
 $\sigma = 0$. It is also possible to  obtain the eigenvalues numerically at each point on a discrete grid $\{\sigma_\imath\}$.
Both of these analytical and numerical approximations can be used to confirm that the eigenvalues are distinct throughout the domain $\sigma \in (0,\sigma_+)$, hence the system is strongly hyperbolic and, equivalently, well-posed \cite{hilditchIntroductionWellposednessFreeevolution2013}. Additionally, both methods  lead to the result:
\begin{equation}
\begin{gathered}
  \lambda _{{\text{out}}}^{|s|}=\lambda _{{\text{out}}}^{|s| + 1} = \ldots =\lambda _{{\text{out}}}^{{\ell _{\max }}} = 0,\quad  \lambda _{{\text{in}}}^{|s|},\lambda _{{\text{in}}}^{|s| + 1}, \ldots ,\lambda _{{\text{in}}}^{{\ell _{\max }}} < 0\quad {\text{at}}\quad \sigma  = {\sigma _{{+ }}}  \\ 
  \lambda _{{\text{in}}}^{|s|}=\lambda _{{\text{in}}}^{|s| + 1} = \ldots =\lambda _{{\text{in}}}^{{\ell _{\max }}} = 0, \quad  \lambda _{{\text{out}}}^{|s|},\lambda _{{\text{out}}}^{|s| + 1}, \ldots ,\lambda _{{\text{out}}}^{{\ell _{\max }}} > 0 \quad {\text{at}}\quad \sigma  =0  \\ 
\end{gathered} 
\end{equation}
Thus, on  the hyperboloidal slices defined by Eq.~\eqref{null_hyper},
there can be no outgoing waves emanating from the horizon or ingoing waves coming from future null infinity. That is, the correct outflow boundary conditions are automatically satisfied and no spurious waves can enter the computational domain
 $\sigma \in [0,  {\sigma _{{+ }}}]$.

\section{Numerical Scheme} \label{num_methods}
\subsection{Method of lines}
\label{matri}
For uncoupled hyperbolic partial differential equations typically encountered in black-hole perturbation theory,  the method of lines  \cite{kreissMethodLinesHyperbolic1992}
can be used to evolve the system. More precisely for a hyperbolic system of partial differential equations 
\begin{equation}\label{eq:GenericHyperbolicPDE}
\partial_{t}u(t,x) = \hat{L}(u(t,x))
\end{equation}
where $\hat L$ is a (possibly nonlinear) spatial differential operator, one proceeds by approximating the field $u(t,x)$ on a discrete spatial grid $\{x_\imath\}_{\imath=0}^N$ so that $u(t,x) \rightarrow \mathbf{u} (t) $. The components $u(t,x_\imath) := u_\imath(t)$ of the vector $\mathbf{u} (t)$ are the values of the field evaluated at the gridpoints. Heuristically, this converts the problem from a partial differential equation in spacetime variables $(t,x)$ to a  system of coupled ordinary differential equations in one time variable $t$,
\begin{equation}\label{eq:GenericODE}
      \frac{d \mathbf{u}}{d t}  =\mathbf{L}(\mathbf{u}).
\end{equation}
where  
the matrix operator $\mathbf{L}$ couples the ordinary differential equations.

If we specify that the differential operator $\hat L$ in Eq.~(\ref{eq:GenericHyperbolicPDE}) is in fact linear, then its spatial discretization is also linear:
\begin{equation}  \label{eq:Ldotu}
   \hat L u(t,x) \rightarrow (\hat L u(t,x) )_\imath = \sum_\jmath {L}_{\imath \jmath} u_\jmath(t).
\end{equation}
That is,  the differential operator $\hat L$, upon discretization, amounts to a matrix $\mathbf{L}$ which then multiplies  a vector $\mathbf{u}(t)$ representing the discrete approximation to the field $u(t,x)$. In matrix product notation, the differential
equation \eqref{eq:GenericODE} becomes
\begin{equation} \label{eq:GenericLinearODE}
    \frac{d \mathbf{u}}{d t} = \mathbf{L} \, \mathbf{u}.
\end{equation}
Written in the form of Eq.~\eqref{eq:GenericLinearODE}, high accuracy solutions to the original partial differential equation can then be obtained using time-symmetric numerical techniques explored in earlier work \cite{markakis_time-symmetry_2019,oboyleConservativeEvolutionBlack2022}.

 Our goal  is thus to be able  to apply the method of lines to
  Eq.~\eqref{time1space2} and write it in the form of Eq.~\eqref{eq:GenericLinearODE}.
The main complication with applying the method of lines to Eq.~\eqref{time1space2} is the coupling between $\ell$--modes.
This coupling means that we need to simultaneously keep track of the value of  all $\ell$--modes (or a sufficient number to ensure accuracy of the solution) of the field during evolution. 
One possibility is to invert the matrix on the left-hand side of 
Eq.~\eqref{time1space2}. As this matrix is dependent on position
$\sigma$, upon spatial discretization, the inversion must be performed at each grid-point
$\sigma_i$. An efficient implementation of this approach,  using the Thomas algorithm to solve for the modes of the scalar wave equation, is described in Ref.\cite{dolan_superradiant_2013}. Even though this point-by-point inversion  is an option for explicit time-stepping schemes, such as Runge-Kutta, it is not an efficient option for  implicit schemes, as it does not bring the system into the form of Eq.~\eqref{eq:GenericLinearODE}.

To achieve our goal, we discretize the black-hole exterior, defined by the compactified radial domain $\sigma \in [0,\sigma_+]$, on a set of  gridpoints
$\{\sigma_\imath\}_{\imath=0}^N$. The  value of the fields
$\varphi$ and $\pi$
 for mode $\ell$ at spatial node $\sigma_\jmath$ will be denoted by 
 $\varphi_{\ell}^\imath(\tau):=\varphi_\ell (\tau,\sigma_\imath)$ and
  $\pi_{\ell}^\imath(\tau):=\pi_\ell (\tau,\sigma_\imath)$ respectively (with the spin index $(s)$ dropped for clarity).
Discretization implies that the spatial derivative operators
$\partial_\sigma$ and $\partial_\sigma^2$ in Eq.~\eqref{time1space2}
will amount to matrices, which can be computed using polynomial collocation (finite-difference or pseudo-spectral) methods\cite{markakisDiscontinuousCollocationMethods2021,oboyleConservativeEvolutionBlack2022}. We  use
 $D_{\imath \jmath}$ and $D^{(2)}_{\imath \jmath}$ to denote the first and second order differentiation matrices respectively: 
\begin{eqnarray*}
    \partial_\sigma u \big|_{\sigma=\sigma_\imath} \simeq ~ \sum_{\jmath=0}^N {D}_{\imath \jmath} u_\jmath, \quad \partial^2_\sigma u \big|_{\sigma=\sigma_\imath} \simeq ~ \sum_{\jmath=0}^N {D}^{(2)}_{\imath \jmath} u_\jmath
\end{eqnarray*}
For instance, a Chebyschev pseudo-spectral method in an interval $\sigma\in[a,b]$, the Chebyschev-Gauss-Lobatto nodes are given by
\begin{equation}  \label{ChebyshevNodes}
    \sigma_\imath=\frac{{{b+a} }}{2} + \frac{{{b} - {a}}}{2} z_\imath, \quad z_\imath = \sin \bigg(\frac{2\imath-N}{2N} \pi\bigg),\quad \imath=0,1,\dots,N
\end{equation}
The above formula for  $z_\imath$ has been preferred over $z_\imath = - \cos(\imath \pi/N)$ because the former yields points $\{ z_\imath\}$ with reflection symmetry about the origin  $z =0$  in floating point arithmetic. The first derivative operator on this grid is
\begin{equation}
    {D}^{}_{\imath \jmath} = 
\frac{2}{b-a}    \begin{cases}
        \frac{c_\imath (-1)^{\imath+\jmath}}{c_\jmath (z_\imath-z_\jmath)} & \imath \neq \jmath\\
        -\frac{z_\jmath}{2(1-z_\jmath^2)} & \imath = \jmath \neq 0,N\\
        -\frac{2 N^2 + 1}{6} & \imath = \jmath = 0\\
        \frac{2 N^2 + 1}{6} & \imath = \jmath = N
    \end{cases}
\end{equation}
where $c_0 = c_N = 2$ and $c_1, \dotsc, c_{N-1} = 1$. 

The second derivative
operator can be evaluated by $ \mathbf{D}^{(2)} = \mathbf{D}^2$
or, equivalently \cite{canuto_2006},
\begin{equation}\label{ChebyshevD2}
D_{\imath \jmath}^{(2)} = {\left( {\frac{2}{{b - a}}} \right)^2}\left\{ {\begin{array}{*{20}{l}}
{\frac{{{{( - 1)}^{\imath + \jmath}}}}{{{c_\jmath}}}\frac{{z_\imath^2 + {z_\imath}{z_\jmath} - 2}}{{(1 - z_\imath ^2){{({z_\imath} - {z_j})}^2}}}}&{\imath \ne \jmath,{\;\;\; }\imath \ne 0,\,{\;\;\; }\imath \ne N}\\
{\frac{2}{3}\frac{{{{( - 1)}^\jmath}}}{{{c_j}}}\frac{{(2{N^2} + 1)(1 + {z_j}) - 6}}{{{{(1 + {z_j})}^2}}}}&{i \ne j,{\;\;\; }\imath = 0}\\
{\frac{2}{3}\frac{{{{( - 1)}^{\jmath + N}}}}{{{c_\jmath}}}\frac{{(2{N^2} + 1)(1 - {z_\jmath}) - 6}}{{{{(1 - {z_\jmath})}^2}}}}&{\imath \ne \jmath,{\;\;\; }\imath = N}\\
{ - \frac{{({N^2} - 1)(1 - z_\jmath^2) + 3}}{{3{{(1 - z_\jmath^2)}^2}}}}&{\imath = \jmath,\,{\;\;\; }\imath \ne 0,{\;\;\; }\imath \ne N}\\
{\frac{{{N^4} - 1}}{{15}}}&{\imath = \jmath = 0{\;\;\; \rm{ or }\;\;\; }N}
\end{array}} \right.
\end{equation}
The Chebyshev differentiation matrices can be constructed automatically via
the Wolfram Language commands:
 \begin{verbatim}
D1=NDSolve`FiniteDifferenceDerivative[Derivative[1],X,
"DifferenceOrder"->"Pseudospectral"]@"DifferentiationMatrix"
\end{verbatim}
 \begin{verbatim}
D2=NDSolve`FiniteDifferenceDerivative[Derivative[2],X,
"DifferenceOrder"->"Pseudospectral"]@"DifferentiationMatrix"
\end{verbatim}
which respectively return the  matrices $\mathbf{D}^{},\mathbf{D}^{(2)}$ for a list $\texttt{X}$ of nodes given by Eq.~\eqref{ChebyshevNodes}. 
One can switch from pseudo-spectral to $n^{\rm th}$-order  finite-difference methods by  simply replacing the nodes \eqref{ChebyshevNodes} with equidistant grid-points and specifying the option\\
\texttt{"DifferenceOrder"->n} \\
for some $n \in \mathbb{N}$ in the commands above. 

For any collocation method, 
Eq.~\eqref{time1space2}
 takes the form of a coupled system of linear ordinary differential equations
\begin{equation} \label{CoupledODESystem}
\sum\limits_{k = |s|}^{{\ell _{\max }}} {\sum\limits_{\jmath =0}^N {\left( {\begin{array}{*{20}{c}}
  {\delta _{\ell k}^{\imath \jmath}}&0 \\ 
  0&{{\text{A}}_{\ell k}^{\imath \jmath}} 
\end{array}} \right)\frac{d}{{d\tau }}\left( {\begin{array}{*{20}{c}}
  {\varphi _k^\jmath} \\ 
  {\pi _k^\jmath} 
\end{array}} \right)} }  = \sum\limits_{k = |s|}^{{\ell _{\max }}} {\sum\limits_{\jmath =0}^N {\left( {\begin{array}{*{20}{c}}
  0&{\delta _{\ell k}^{\imath \jmath}} \\ 
  {{\rm M}_{\ell k}^{\imath \jmath}}&{{\rm N}_{\ell k}^{\imath \jmath}} 
\end{array}} \right)\left( {\begin{array}{*{20}{c}}
  {\varphi _k^\jmath} \\ 
  {\pi _k^\jmath} 
\end{array}} \right)} } 
\end{equation}
that can be evolved in time $\tau$.
Here, the indices $\imath, \jmath =0,1,\dots,N$ label  different gridpoints and the indices
$k, \ell=|s|, \dots ,\ell_{\max}  $  label  different $\ell$--modes. The rank-4 array ${\delta _{\ell k}^{\imath \jmath}}:={\delta _{\ell k} \delta^{\imath \jmath}} $ is the generalized Kronecker delta and 
\begin{equation} \label{rank4arrays}
\begin{aligned}
 & {\text{A}}_{\ell k}^{\imath \jmath} := {\delta _{\imath \jmath}}{{\rm A}_{\ell k}}({\sigma _\imath}) \\ 
&  {\rm M}_{\ell k}^{\imath \jmath} := {\Delta _{\ell k}}({\sigma _\imath}){D_{\imath \jmath}} + {{\rm E}_{\ell k}}({\sigma _\imath})D_{\imath \jmath}^{(2)} + {\delta _{\imath \jmath}}{{\rm Z}_{\ell k}}({\sigma _\imath}) \\ 
&  {\rm N}_{\ell k}^{\imath \jmath} := {{\rm B}_{\ell k}}({\sigma _\imath}){\delta _{\imath \jmath}} + {\Gamma _{\ell k}}({\sigma _\imath}){D_{\imath \jmath}} \\ 
\end{aligned} 
\end{equation}
are rank-4 arrays defined by evaluating 
Eqs.~\eqref{eq:2} on the spatial grid $\{\sigma_\imath \}$.
No summation over repeated indices is implied in the above expressions. The array products above are a combination of Hadamard products (also known as element-wise or Schur products) and Kronecker products, and can thus be performed efficiently using Intel Math Kernel  Libraries (MKL). In Wolfram language, the commands\\ \texttt{A*B} 
\\or\\
 \texttt{KroneckerProduct[A,B]} \\
 respectively compute  the Hadamard or Kronecker products of the matrices \texttt{A} and \texttt{B} via suitable Intel MKL function calls, and the arrays \eqref{rank4arrays}  are stored in memory.

\subsection{Matricization}
With the above spatial discretization, 
 the vector of state variables 
\begin{equation}
u_\ell (\tau,\sigma) = \left( {\begin{array}{*{20}{c}}
  {\varphi _\ell(\tau,\sigma)} \\ 
  {\pi _\ell(\tau,\sigma) }
\end{array}} \right), \quad \ell=|s|, \dots, \ell_{\max}
\end{equation}
which appears in  
Eq.~\eqref{time1space2}  will amount to a rectangular matrix $\mathbf{U} (\tau) $ of dimensions $(2N+2)\times (\ell_{\max}-|s|)$ with elements
\begin{equation}
U_{i \ell} (\tau):= u_\ell (\tau,\sigma_i) =
 \left( {\begin{array}{*{20}{c}}
  {\varphi _\ell(\tau,\sigma_i)} \\ 
  {\pi _\ell(\tau,\sigma_{i-N+1}) }
\end{array}} \right)
= \left( {\begin{array}{*{20}{c}}
  {\varphi _\ell^i(\tau)} \\ 
  {\pi _\ell^{i-N+1}(\tau) }
\end{array}} \right), \quad i=0,1,\dots,2N+2  
\end{equation}
So, for $i=0, \dots N$, $U_{i \ell}=\varphi_{\ell}^i$ will be the value of the field
$\varphi$
 for mode $\ell$ at spatial node $\sigma_i$ and, for 
$i= N+1, \dots ,2N+2$, $U_{i \ell}=\pi_{\ell}^{i-N+1}$ will be the value of the variable
$\pi$
 for mode $\ell$ at spatial node $\sigma_{i-N+1}$. To 
bring the system to the desired form \eqref{eq:GenericLinearODE}, we use vectorisation (or flattening) \cite{landsbergTensorsGeometryApplications2012}, which is a linear transformation of a $m\times n$ matrix to a $mn$ column vector, to  write the rectangular matrix  $\mathbf{U}  $ as a column vector
$\mathbf{u} = {\rm{vec }} \; \mathbf{U}$.
For instance,  if $|s|=0$, $N=2$ and $\ell_{\max}=1$, the below example shows how  a $(\ell_{\max}-|s|+1)\times (N+1)=2\times 3 $ matrix  is flattened to a column vector of length $(\ell_{\max}-|s|+1) (N+1)=6$:
\begin{equation}
\mathbf{U}=\left( {\begin{array}{*{20}{c}}{{U_{00}}}&{{U_{10}}}\\{{U_{01}}}&{{U_{11}}}\\{{U_{02}}}&{{U_{12}}}\end{array}} \right) \rightarrow \mathbf{u}= \left( {\begin{array}{*{20}{c}}{{U_{00}}}\\{{U_{01}}}\\{{U_{02}}}\\{{U_{10}}}\\{{U_{11}}}\\{{U_{12}}}\end{array}} \right).
\end{equation}
Furthermore, the rank-4 arrays $\mathbf{A}, \mathbf{M}, \mathbf{N}$ defined in Eq.~\eqref{rank4arrays} are matricized (or flattened) to square matrices
$\mathbb{A}, \mathbb{M}, \mathbb{N}$
 \cite{landsbergTensorsGeometryApplications2012}. 
For instance, if $|s|=0$, $N=2$ and $\ell_{\max}=1$, the  array
$\mathbf{A}$, with components ${{A}}^{\imath \jmath}_{\ell k}$, 
is flattened to a square matrix $\mathbb{A} $ as follows:
\begin{equation}
 \begin{split}
\mathbf{A} = \left( {\begin{array}{*{20}{c}}{\left( {\begin{array}{*{20}{c}}{A_{00}^{00}}&{A_{01}^{00}}\\{A_{10}^{00}}&{A_{11}^{00}}\end{array}} \right)}&{\left( {\begin{array}{*{20}{c}}{A_{00}^{01}}&{A_{01}^{01}}\\{A_{10}^{01}}&{A_{11}^{01}}\end{array}} \right)}&{\left( {\begin{array}{*{20}{c}}{A_{00}^{02}}&{A_{01}^{02}}\\{A_{10}^{02}}&{A_{11}^{02}}\end{array}} \right)}\\{\left( {\begin{array}{*{20}{c}}{A_{00}^{10}}&{A_{01}^{10}}\\{A_{10}^{10}}&{A_{11}^{10}}\end{array}} \right)}&{\left( {\begin{array}{*{20}{c}}{A_{00}^{11}}&{A_{01}^{11}}\\{A_{10}^{11}}&{A_{11}^{11}}\end{array}} \right)}&{\left( {\begin{array}{*{20}{c}}{A_{00}^{12}}&{A_{01}^{12}}\\{A_{10}^{12}}&{A_{11}^{12}}\end{array}} \right)}\\{\left( {\begin{array}{*{20}{c}}{A_{00}^{20}}&{A_{01}^{20}}\\{A_{10}^{20}}&{A_{11}^{20}}\end{array}} \right)}&{\left( {\begin{array}{*{20}{c}}{A_{00}^{21}}&{A_{01}^{21}}\\{A_{10}^{21}}&{A_{11}^{21}}\end{array}} \right)}&{\left( {\begin{array}{*{20}{c}}{A_{00}^{22}}&{A_{01}^{22}}\\{A_{10}^{22}}&{A_{11}^{22}}\end{array}} \right)}\end{array}} \right) \rightarrow\\
\mathbb{A}=
\left( {\begin{array}{*{20}{c}}{A_{00}^{00}}&{A_{01}^{00}}&{A_{00}^{01}}&{A_{01}^{01}}&{A_{00}^{02}}&{A_{01}^{02}}\\{A_{10}^{00}}&{A_{11}^{00}}&{A_{10}^{01}}&{A_{11}^{01}}&{A_{10}^{02}}&{A_{11}^{02}}\\{A_{00}^{10}}&{A_{01}^{10}}&{A_{00}^{11}}&{A_{01}^{11}}&{A_{00}^{12}}&{A_{01}^{12}}\\{A_{10}^{10}}&{A_{11}^{10}}&{A_{10}^{11}}&{A_{11}^{11}}&{A_{10}^{12}}&{A_{11}^{12}}\\{A_{00}^{20}}&{A_{01}^{20}}&{A_{00}^{21}}&{A_{01}^{21}}&{A_{00}^{22}}&{A_{01}^{22}}\\{A_{10}^{20}}&{A_{11}^{20}}&{A_{10}^{21}}&{A_{11}^{21}}&{A_{10}^{22}}&{A_{11}^{22}}\end{array}} \right)
\end{split}
\end{equation}
and similarly for the the  arrays   $\mathbf{M} \rightarrow \mathbb{M}$ and
 $\mathbf{N} \rightarrow \mathbb{N}$.
After this matricization, the coupled  system \eqref{CoupledODESystem} takes the form 
\begin{equation}
\left( {\begin{array}{*{20}{c}}
  \mathbb{I}&0 \\ 
  0&\mathbb{A}\ 
\end{array}} \right)\frac{{d{\mathbf{u}}}}{{d\tau }} = \left( {\begin{array}{*{20}{c}}
  0&\mathbb{I}\ \\ 
  \mathbb{M} & \mathbb{N} 
\end{array}} \right){\mathbf{u}}
\end{equation}
%
where $\mathbb{I}$ is a $
[(N+1) ({\ell_{\max}-|s|+1})] \times [(N+1) ({\ell_{\max}-|s|+1})]  $
 unit matrix and 0 is the null matrix of the same dimensions.
As the matrix $\mathbb{A}$ is square and well-conditioned, we can typically compute its matrix inverse  $\mathbb{A}^{-1}$. Moreover, since this matrix is time-independent, its inverse can be computed in advance for a given grid and stored in memory. This  allows us to finally write  the  coupled  system \eqref{CoupledODESystem}   in the form 
\eqref{eq:GenericLinearODE}, that is,
\begin{equation} \label{dudtLu}
\frac{{d{\mathbf{u}}}}{{d\tau }} = \mathbf{L}  \,{\mathbf{u}}
\end{equation}
where
%
\begin{equation} \label{Lmatrix}
\mathbf{L} = \left( {\begin{array}{*{20}{c}}0&\mathbb{I} \\\mathbb{A}^{-1}  \mathbb{M} & \mathbb{A}^{-1} \mathbb{N}  \end{array}} \right)
\end{equation}
is a square matrix of size  $
[(2N+2) ({\ell_{\max}-|s|+1})] \times [(2N+2) ({\ell_{\max}-|s|+1})]  $ which is constant for a given grid.

\subsection{Time-symmetric evolution}
The matricization approach drastically simplifies  numerical evolution. Once the matrix $\mathbf{L}$ is constructed by means of Eqs.~\eqref{Lmatrix}
and \eqref{rank4arrays} and stored in memory, evolution amounts to a simple matrix multiplication. In fact, due to
the linearity of Eq.~\eqref{dudtLu}, an analytical solution in time
can be written using the matrix exponential:
\begin{equation} \label{ueLtu}
{\mathbf{u}}(\tau) = e^ {\mathbf{L} \,\tau}  {\mathbf{u}(0)}
\end{equation}
which can be evaluated using exponential integrators \cite{hochbruckExponentialIntegrators2010} for any time $\tau$.
This side-steps the need for numerical integration in time, which has typically been performed via explicit methods
in the
literature.

In the absence of efficient numerical matrix exponential libraries, it can be  faster and more accurate to apply a time-symmetric integration method, such as Hermite integration, to evaluate
the integral \begin{equation}   \label{eq:GenericIntEqLu}
    \mathbf{u}_{\nu+1} = \mathbf{u}_\nu + \mathbf{L}  \int_{\tau_\nu}^{\tau_{\nu+1}}
\mathbf{u}(\tau)~ d\tau,
\end{equation}
where $ \mathbf{u}(\tau_\nu) := \mathbf{u}_\nu $ and  $ \mathbf{u}(\tau_{\nu+1}) := \mathbf{u}_{\nu+1} $, for small time-steps $\Delta \tau=\tau_{\nu+1}-\tau_\nu$.  
Equivalently, 
one may write Eq.~\eqref{ueLtu}  as
\begin{equation}  \label{eq:GenericExpdtLu}
  \mathbf{u}_{\nu+1} = e^{ \mathbf{L}   \,\Delta \tau}  \mathbf{u}_\nu=\mathbf{u}_\nu+ ( e^{ \mathbf{L}   \,\Delta \tau}- \mathbf{I})  \mathbf{u}_\nu,
\end{equation}
where $\mathbf{I}$ is the unit matrix of the same size as $\mathbf{L}$, and use a series expansion to approximate the matrix exponential.

For a linear system, applying a Hermite integration method of order $p$ to evolve Eq.~\eqref{dudtLu} is equivalent to applying a \textit{symmetric 2-point  Taylor expansion} around  $\tau_\nu$ and $\tau_{\nu+1}$ to integrate Eq.~\eqref{eq:GenericIntEqLu} or applying a \textit{symmetric Pad\'e  expansion} to 
approximate the matrix exponential in Eq.~\eqref{eq:GenericExpdtLu}.
The symmetric  Pad\'e  approximant of order $p$ is
\begin{equation} \label{Pade}
    e^{\Delta t \, \mathbf{L}} \simeq  \bigg[ \sum_{q = 0}^p c_{p q} ~(-  \mathbf{L} \, \Delta \tau)^{q}\bigg]^{-1}  {\sum_{p = 0}^p c_{pq} ~( \mathbf{L} \, \Delta \tau)^{q}} \, + \frac{(p!)^2}{(2p+1)!(2p)!} \mathcal{O}(\Delta \tau^{2p+1})  ,
\end{equation}
where
\begin{equation} \label{eq:clm}
    c_{pq} = \frac{p!(2p-q)!}{q!(2p)!(p-q)!}
\end{equation}
The advantages of time-symmetric integration  have been detailed in earlier work \cite{markakis_time-symmetry_2019,oboyleConservativeEvolutionBlack2022}.
The  symmetry of Eq.~\eqref{eq:GenericIntEqLu} under the exchange $\tau_\nu \leftrightarrow \tau_{\nu+1}$  is preserved by the  approximant \eqref{Pade}
and, for linear PDEs arising from Hamiltonian systems, has been shown to
preserve energy and phase-space volume.
Moreover, by construction, the spectral radius of the  matrix \eqref{Pade} is
$\rho=1$ for any spatial discretization that satisfies appropriate boundary conditions, regardless of the time-step $\Delta \tau$. (This can be confirmed by computing the maximum absolute eigenvalue of the matrix   \eqref{Pade}
for a given collocation method.) This class of time-symmetric schemes is therefore \textit{unconditionally stable} and there is no time-step restriction  \cite{courantPartialDifferenceEquations1967}).

In contrast,
for a linear system, applying a Runge-Kutta method of order $p$ to evolve Eq.~\eqref{dudtLu} would be equivalent to applying a \textit{1-point Taylor expansion} around  $\tau_n$ to integrate Eq.~\eqref{eq:GenericIntEqLu} or to expand the matrix exponential in Eq.~\eqref{eq:GenericExpdtLu}:
\begin{equation} \label{Taylor}
    e^{\Delta \tau \, \mathbf{L}} \simeq 
\sum_{q=0}^p \frac{( \mathbf{L}  \,\Delta \tau )^q}{q!} + \frac{{{1{}}}}{{(p+1){{!}}}}  \mathcal{O} (\Delta
{\tau^{p+1}}).
\end{equation} 
Such approximants are not symmetric under the exchange $\tau_\nu \leftrightarrow \tau_{\nu+1}$. Explicit schemes that violate time-symmetry, such as Runge-Kutta,  are conditionally stable. That is, the spectral radius of the matrix \eqref{Taylor}
exceeds unity if the time-step $\Delta \tau$ exceeds the Courant limit  \cite{courantPartialDifferenceEquations1967}). (This can be confirmed by computing the maximum absolute eigenvalue of the matrix  \eqref{Taylor}
for a given collocation method.) Thus, one must use more/smaller time-steps, making a simulation more expensive computationally. In addition, such schemes  typically fail to numerically conserve constants of the system. Finally, the error term in Eq.~\eqref{Pade}
falls off much faster than the error term in  \eqref{Taylor} for increasing order $p$.
Hence, for long-time integration, as required e.g. by  the applications outlined in the introduction, 
time-symmetric schemes such as Eq.~\eqref{Pade}
 are preferable.

For $p=2$, Eq.~\eqref{Pade} yields a scheme which amounts to applying the trapezium rule to Eq.~\eqref{eq:GenericIntEqLu} (and is equivalent to the
Crank-Nicolson scheme if accompanied by second order finite-differencing in space):
\begin{subequations}
\begin{eqnarray}  \label{eq:LD2OperatorImprecise}
{\mathbf{u}}{ _{\nu + 1}} \mkern-10mu  &=& \mkern-12mu{\bigg({\mathbf{I}} - \frac{{\mathbf{L}\Delta \tau}}{2}{}\bigg)^{ - 1}}  \bigg({\mathbf{I}} + \frac{{\mathbf{L}\Delta \tau}}{2}{\mathbf{}}\bigg)   {\mathbf{u}}{_\nu} 
  \\  \label{eq:LD2OperatorPrecise}
     &=&  \mkern-12mu \mathbf{u}_\nu +\bigg( \mathbf{I} - \frac{\mathbf{L}\Delta \tau}{2} \mathbf{} \bigg)^{-1}  \mathbf{L} \,\Delta \tau   
\;\mathbf{u}_\nu .
\end{eqnarray}
\end{subequations}
Similarly, for $p=4$,  Eq.~\eqref{Pade} yields a scheme equivalent to applying the  Hermite integration rule to Eq.~\eqref{eq:GenericIntEqLu}:
\begin{subequations}
\begin{eqnarray}  \label{eq:LD4OperatorImprecise}
{\mathbf{u}}{ _{\nu + 1}} \mkern-10mu & =&  \mkern-10mu {\left[ {{\mathbf{I}} - \frac{{\mathbf{L}\Delta \tau }}{2}{\mathbf{}}   \left( {{\mathbf{I}} - \frac{{\mathbf{L}\Delta \tau }}{6}{\mathbf{}}} \right)} \right]^{ - 1}}  \left[ {{\mathbf{I}} + \frac{{\mathbf{L}\Delta \tau }}{2}{\mathbf{}}  \left( {{\mathbf{I}} + \frac{{\mathbf{L}\Delta \tau }}{6}{\mathbf{}}} \right)} \right]  {\mathbf{u}}{_\nu}   \quad \quad\quad
  \\  \label{eq:LD4OperatorPrecise}
    & =& \mkern-12mu  \mathbf{u}_\nu + \bigg[ \mathbf{I} - \frac{\mathbf{L}\Delta \tau}{2} \mathbf{}  \bigg(\mathbf{I}-\frac{\mathbf{L}\Delta \tau}{6} \mathbf{}  \bigg) \bigg]^{-1}     \mathbf{L} \,\Delta \tau \;
\mathbf{u}_\nu 
\end{eqnarray}
\end{subequations}
For $p=6$,  Eq.~\eqref{Pade} yields a scheme equivalent to applying Lotkin's integration rule to Eq.~\eqref{eq:GenericIntEqLu}:
\begin{subequations}
\begin{eqnarray}  \label{eq:LD6OperatorImprecise}
{\mathbf{u}}{_{\nu + 1}} \mkern-10mu & =&  \mkern-10mu 
{\left[ {{\mathbf{I}} - \frac{{\mathbf{L}\Delta \tau }}{2}  \left( {{\mathbf{I}} - \frac{{\mathbf{L}\Delta \tau }}{5}  \left( {{\mathbf{I}} - \frac{{{\mathbf{L}}\Delta \tau }}{{{\text{12}}}}} \right)} \right)} \right]^{ - 1}} 
\nonumber    
 {\left[ {{\mathbf{I}} + \frac{{\mathbf{L}\Delta \tau }}{2}  \left( {{\mathbf{I}} + \frac{{\mathbf{L}\Delta \tau }}{5}  \left( {{\mathbf{I}} + \frac{{{\mathbf{L}}\Delta \tau }}{{{\text{12}}}}} \right)} \right)} \right]} 
  {\mathbf{u}}{_\nu}   \quad 
  \\  \label{eq:LD6OperatorPrecise}
  \mkern-10mu &  =& \mkern-10mu \mathbf{u}_\nu 
        +{\left[ {{\mathbf{I}} - \frac{{\mathbf{L}\Delta \tau }}{2} \mkern-5mu \left( {{\mathbf{I}} - \frac{{\mathbf{L}\Delta \tau }}{5} \mkern-5mu \left( {{\mathbf{I}} - \frac{{{\mathbf{L}}\Delta \tau }}{{{\text{12}}}}} \right)} \right)} \right]^{ - 1}} {\mathbf{L}}\Delta \tau \left( {{\mathbf{I}} + \frac{{\mathbf{L}\Delta \tau }}{{60}} } \right)   \mathbf{u}(\tau_\nu) \quad
\end{eqnarray}
\end{subequations}
%
%
In the above examples, terms polynomial in $\Delta \tau$ have been factored in Horner form to improve arithmetic precision. Crucially, instead of the  matrix multiplications~\ref{eq:LD2OperatorImprecise}, \ref{eq:LD4OperatorImprecise}
and \ref{eq:LD6OperatorImprecise} which entail high round-off error accumulation (due to non-compensated summations in each matrix multiplication) we have  separated each time-step into a matrix multiplication and addition, as dictated by Eqs.~\ref{eq:LD2OperatorPrecise}, \ref{eq:LD4OperatorPrecise}
and \ref{eq:LD6OperatorPrecise}. Prior work   \cite{markakis_time-symmetry_2019,oboyleConservativeEvolutionBlack2022} has shown that separating the additive term substantially reduces the round-off error accumulated in each time-step and is essential for conserving energy and phase-space volume numerically. Such errors could be further reduced by matrix multiplication libraries that implement compensated summation. 
 Because generalized matrix multiply and addition operations are parallelized and heavily optimized by modern CPU and GPU libraries, this scheme can be implemented  efficiently without heavy programming effort. 
Due to their  conservation properties,  low truncation and round-off errors, and linear algebraic library support, the above time-symmetric schemes are ideal for long time numerical evolution in black hole perturbation theory.

\section{Numerical Evolution}
\label{}

\subsection{Late-time Tails}
\label{price_tails}
Price's   law \cite{priceNonsphericalPerturbationsRelativistic1972a,priceNonsphericalPerturbationsRelativistic1972b}  may be heuristically stated as follows: any radiatable inhomogeneities in the spacetime geometry outside a black hole will be radiated away. Such inhomogeneities can be quantified as nonzero higher ($\ell \ge |s|$) multipole modes of Schwarzschild spacetime perturbations.  A consequence of Price's theorem is that higher order modes will decay polynomially in time
\begin{equation}\label{price_tail_rule}
\varphi_{\ell}\propto \tau^{-\Gamma_\ell}
\end{equation} 
where the effective local power index  $\Gamma_\ell$ can be calculated for late times $\tau$ from

\begin{equation}\label{price_tail_formula}
        \Gamma_\ell = \tau \; \frac{\Re(\varphi_\ell) \;\Re(\pi_\ell)+\Im(\varphi_\ell)\;\Im(\pi_\ell)}{\Re(\varphi_\ell)^2+\Im(\varphi_{\ell})^2}.
\end{equation}
where $\Re$ and $\Im$  denote the real and imaginary part of the field respectively. Price's law has been generalized and rigorously proven for Kerr black holes by Angelopoulos et al. \cite{angelopoulosLatetimeTailsMode2021}.

\begin{figure}[ht]
                \centering
                \includegraphics[scale=0.4]{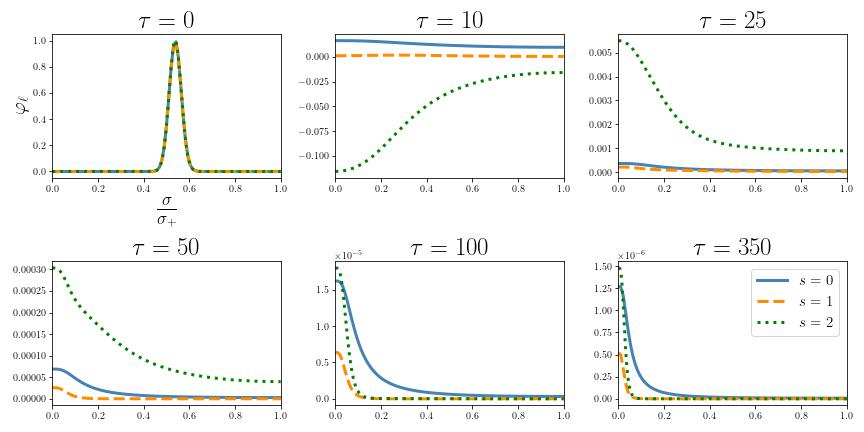}
                \caption{Initial profile of perturbation  $\varphi_{\ell}(\tau,\sigma)$ at $\tau=0$ followed by snapshots of the evolution  up until time $\tau=350$ for multipole mode $\ell = |s|$ and spins $s=0,1,2$.}
        \label{fig:initial_condition}
\end{figure}

Computing  late-time tails numerically requires high  accuracy. Using our methodology to compute tails and confirm that our results align with previously computed tails \cite{raczNumericalInvestigationLatetime2011,zenginogluIntermediateBehaviorKerr2014, csukasNumericalInvestigationDynamics2019} serves as a stringent test to confirm our code and methodology works as intended.
Here, we compute late-time tails  for three different  spin weights: $s=0$, $s=1$ and $s=2$. To produce initial data of compact support away from the horizon, we  perturb the lowest mode $\ell=|s|$ of the field using a Gaussian pulse
\begin{equation}\label{initial_perturb}
   f(\sigma) =       e^{-256(\sigma-1)^2}
\end{equation}
as initial perturbation and  set $\varphi_\ell(\tau,\sigma)|_{\tau=0}=f(\sigma)$ and $\pi_\ell(\tau,\sigma)|_{\tau=0}=0$.  The  initial data and subsequent evolution is shown  in Fig.~\ref{fig:initial_condition}.
As the lowest mode of the field will be set by the spin of the field, for $s=0$ we perturbed the mode $\ell=0$, similarly for $s=1$ we perturbed the $\ell=1$ mode and for $s=2$ we perturbed the $\ell=2$ mode.

In our sub-extremal computations we set the Kerr parameter  to $\chi = \frac{1}{4}$ or $\frac{a}{M}=\frac{1}{2}$. Additionally, because late-time tails decay very rapidly to values below machine $\epsilon$, we truncate the multipole expansion at $\ell_{\max} = 4$ and have focused on axially symmetric modes,  $m=0$, in this section. Increasing $\ell_{\max}$ affects the early quasi-normal ringdown phase, where many $\ell$-multipole modes contribute, but not the late-time tail phase.  (Non-axisymmetric modes are examined in the next section, mainly in the extremal Kerr limit.) We have varied between using pseudo-spectral  and finite-difference methods for discretising the spatial grid. We prefer to use a pseudo-spectral grid when possible because it allows us to calculate accurate solutions with a smaller grid so is faster to run. However there are scenarios (when we try to compute tails away from $\mathscr{I}^{+}$ for $s=1$ and $s=2$) where we do not have sufficient accuracy with a pseudo-spectral grid at late times, as we cannot make the pseudo-spectral grid large enough\footnote{Chebyschev differentiation matrices have a high condition number which increases as $N^2$ or $N^4$ for first and second order derivatives respectively, and matrix multiplications involve summations over elements that are not compensated by typical numerical linear algebra libraries; therefore round-off error exceeds truncation error for  a high number of points.} to resolve sharp features. As shown in  Fig.~\ref{fig:initial_condition}, because of the slower decay rate at $\mathscr{I}^+$, the solution looks more and more like a step function at late times. As demonstrated in Ref.~\cite{zenginogluHyperboloidalStudyTail2008}, resolving this steep gradient requires more grid points near  $\mathscr{I}^+$. If this important, one can use analytical mesh refinement, via a coordinate transformation that stretches the  region near   $\mathscr{I}^+$, to resolve this issue. But this is beyond the scope of  this paper and may not generalize straightforwardly to discontinuous collocation methods\cite{markakisDiscontinuousCollocationMethods2021}, which is a natural next step towards including a point particle source.
Instead, for these scenarios, we use a finite-difference grid with more points to gain the accuracy  needed. One of the advantages of formulating numerical schemes and numerical code in terms of matrices is that the type of discretisation used is effectively a code parameter inputted into a function. This automates any desired  switching between a pseudo-spectral  and a finite-difference grid, as this   requires just  changing the nodes and the differentiation matrices, while the remaining code remains  unaltered.
Our results for $s=0,1,2$ can be seen below in figures \ref{fig:s0}, \ref{fig:s1}, \ref{fig:s2}. 

All results presented here use standard machine arithmetic (FP64 precision) and simulations take only a few minutes    on a 28 Core Intel Xeon Platinum 8173M  CPU, as Intel Math Kernel Libraries utilize all available cores. This is another reason why a matricization is computationally beneficial. As mentioned earlier, because each time-step of the scheme~\eqref{eq:GenericExpdtLu} 
amounts to a simple matrix multiplication and addition, our code can also run on GPUs with FP64 support with minimal changes. In Wolfram Language,  this amounts to simply changing the command \texttt{Dot[ ]} (which uses Intel MKL) to \texttt{CUDADot[ ]} (which uses CuBLAS),  transferring  arrays from host memory to device  memory to accelerate matrix multiply and addition operations, and returning the result to the host in the end of a simulation.  On a NVidia GV100 or a Titan V GPU this results in a speedup by a factor of $\sim 5$. We estimate the speedup to be $\sim 12.5$ on a Nvidia A100 and  $\sim 50$ on a NVidia H100 GPU. However, we find that large matrix multiplications on the GPU performed in FP64 (or \texttt{double}) precision using CuBLAS entail a loss of accuracy of about 2 digits compared to multiplications on the CPU, which are internally performed in  FP80 (or \texttt{long double}) precision using Intel MKL, and these errors may accumulate over a long time evolution. Because of this, accelerating computations on GPUs is more useful when using a very large number of grid-points with sparse finite-differencing arrays.   Similar implementations in other languages such as Julia or Python are straighforward. In Python, one can use external libraries such as NumPy or PyTorch to perform the matrix multiplications on CPUs or GPUs without needing to alter the rest of the code. For spins $s=-1$ or $s=-2$, tails decay too rapidly and reach machine $\epsilon$ before they can be extracted. Thus, FP128 or higher precision is typically required for these spins, which is emulated in software in most CPU architectures. As software arithmetic is orders of magnitude slower, these cases will be revisited in future work. Nevertheless, such precision would only be needed for the purpose of computing late-time tails. For the purpose of computing EMRI waveforms, FP80 or FP64 precision is adequate.
\newpage
\vfill
\begin{figure}[ht]
        \centering
        \begin{subfigure}{0.45\textwidth}
                \centering
                \includegraphics[width=\textwidth]{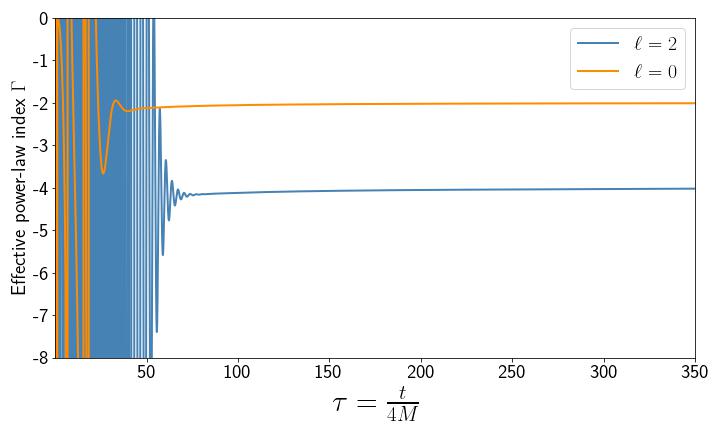}
                \caption{Tails for $(m,s)=(0,0)$ at future null infinity $\mathscr{I}^{+}$, extracted at $\sigma=0$.}
        \end{subfigure}
        \begin{subfigure}{0.45\textwidth}
                \centering
                \includegraphics[width=\textwidth]{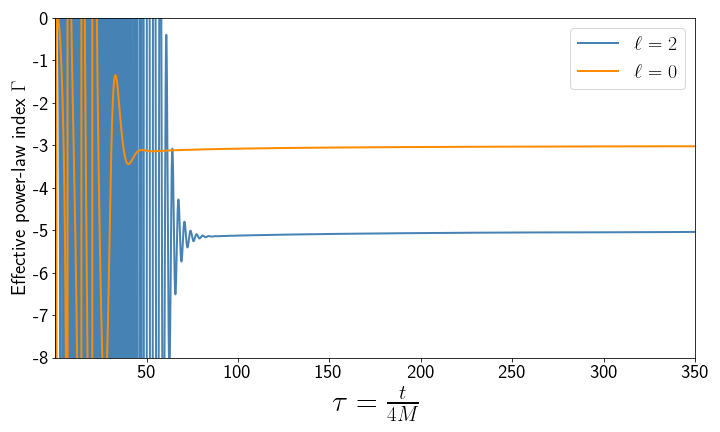}
                \caption{Tails for $(m,s)=(0,0)$ in the computational domain $\sigma \in (0,\sigma_+)$.}
        \end{subfigure}
         \begin{subfigure}{0.45\textwidth}
                        \centering
                        \includegraphics[width=\textwidth]{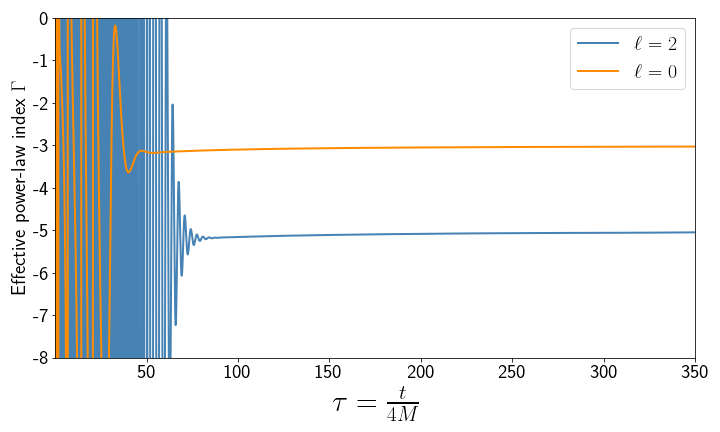}
                        \caption{Tails for $(m,s)=(0,0)$ on the future  horizon $\mathscr{H}^{+}$,  $\sigma=\sigma_+$.}
        \end{subfigure}
        
        \caption{Late-time tails of axisymmetric ($m=0$) scalar ($s=0$) perturbations on Kerr background, computed using a Chebyschev pseudo-spectral spatial grid of $96$ points. For $s=0$, coupling only impacts every second mode; thus, as we perturbed the $\ell=0$ mode, we only have tails for even modes. The Kerr spin parameter is $\frac{a}{M}=\frac{1}{2}$.}
        \label{fig:s0}
\end{figure}
\newpage
\vfill
\begin{figure}[ht]
        \centering
        \begin{subfigure}{0.45\textwidth}
                \centering
                \includegraphics[width=\textwidth]{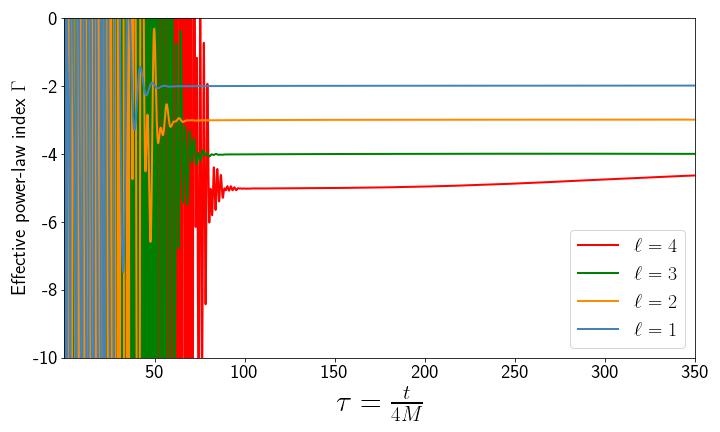}
                \caption{Tails for $(m,s)=(0,1)$ at future null infinity $\mathscr{I}^{+}$, extracted at $\sigma=0$.}
        \end{subfigure}
        \begin{subfigure}{0.45\textwidth}
                \centering
                \includegraphics[width=\textwidth]{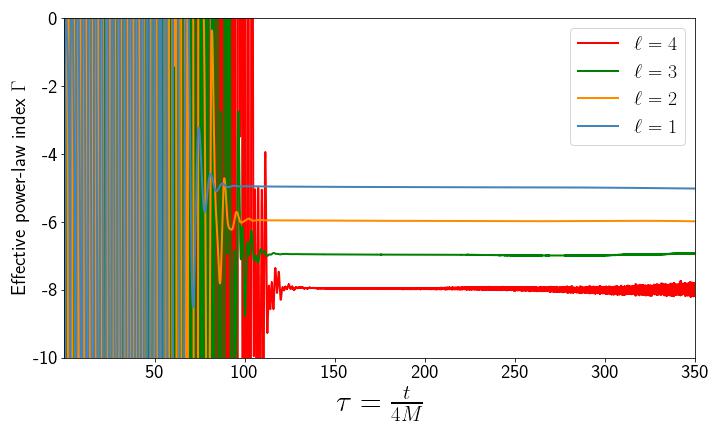}
                \caption{Tails for $(m,s)=(0,1)$ in the computational domain $\sigma \in (0,\sigma_+)$.}
        \end{subfigure}
            \begin{subfigure}{0.45\textwidth}
                \centering
                \includegraphics[width=\textwidth]{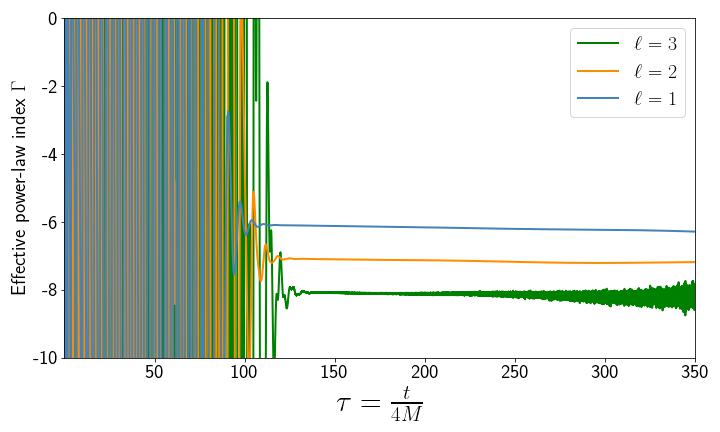}
                \caption{Tails for $(m,s)=(0,1)$ on the future horizon $\mathscr{H}^{+}$,  $\sigma=\sigma_+$.}
            \end{subfigure}    
        \caption{Late-time tails from axisymmetric ($m=0$) perturbations of a  $s=1$ field on a Kerr background, extracted using two different collocation methods. A Chebyschev pseudo-spectral spatial grid of $96$ points yields better results when extracting  tails on future null infinity $\mathscr{I}^{+}$.  For extracting tails within the domain and on the future horizon  $\mathscr{H}^{+}$, a finite-difference grid of $656$ points was used. The Kerr spin parameter is $\frac{a}{M}=\frac{1}{2}$.}
        \label{fig:s1}
\end{figure}
\vfill
\begin{figure}[ht]
        \centering
        \begin{subfigure}{0.45\textwidth}
                \centering
                \includegraphics[width=\textwidth]{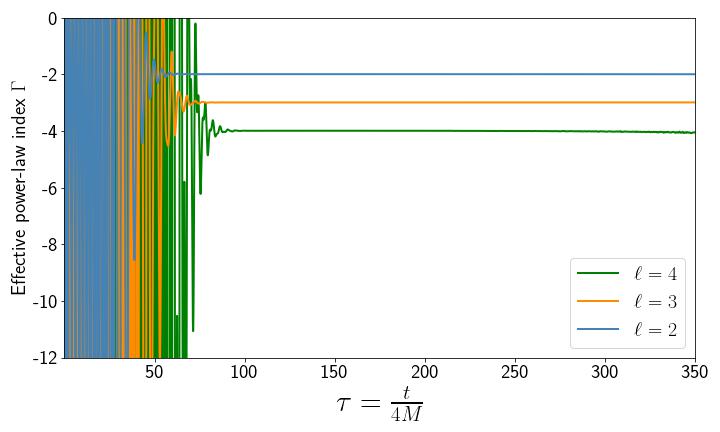}
                \caption{Tails for $(m,s)=(0,2)$ at future null infinity $\mathscr{I}^{+}$, extracted at $\sigma=0$.}
        \end{subfigure}
        \begin{subfigure}{0.45\textwidth}
                \centering
                \includegraphics[width=\textwidth]{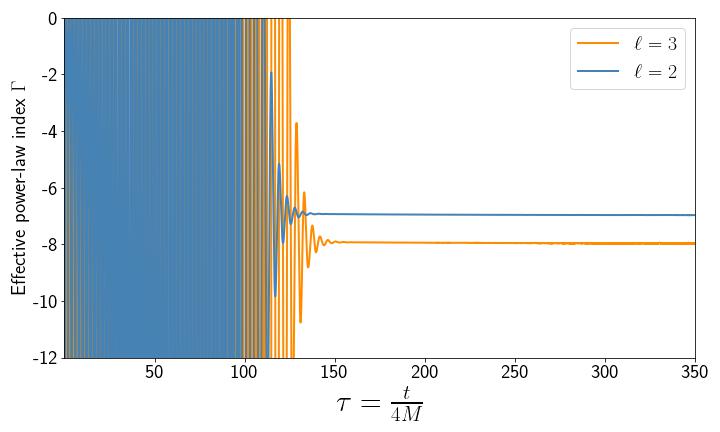}
                \caption{Tails for $(m,s)=(0,2)$ inside the domain  $\sigma \in (0,\sigma_+)$.}
        \end{subfigure}
            \begin{subfigure}{0.45\textwidth}
                \centering
                \includegraphics[width=\textwidth]{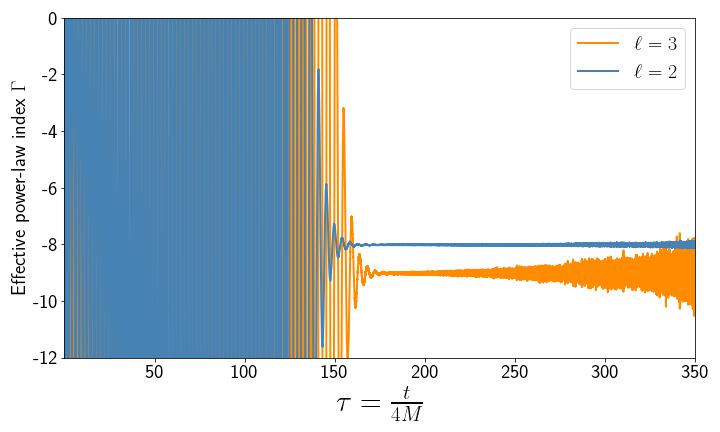}
                \caption{Tails for $(m,s)=(0,2)$ on the horizon $\mathscr{H}^{+}$,  at $\sigma=\sigma_+$.}
            \end{subfigure}
        \caption{Late-time tails from axisymmetric ($m=0$) perturbations on Kerr background, for $s=2$ field. Computed using a pseudo-spectral spatial grid of $316$ points for tails on future null infinity $\mathscr{I}^{+}$, while within the domain and on the future horizon  $\mathscr{H}^{+}$ a finite difference grid of 2,256 points was used. The Kerr spin parameter is $\frac{a}{M}=\frac{1}{2}$.}
        \label{fig:s2}
\label{fig:SchroConvergence}
\end{figure}

\subsection{Aretakis Instability}
\label{aretakis_instability}

Extremal Kerr  $|a|=M$ black holes are subject to an instability discovered by  Aretakis  \cite{aretakisHorizonInstabilityExtremal2013,aretakisDynamicsExtremalBlack2018}. Aretakis' theorem states that, for generic solutions of the wave equation on an extremal Kerr background, the first derivatives fail to decay along the event horizon as advanced time tends to infinity, and moreover, the second derivatives blow up at a polynomial rate.
This  is in sharp contrast to  subextremal  $|a|<M$ Kerr black holes, where it has been shown
\cite{dafermosDecaySolutionsWave2014}  
 that arbitrary derivatives of solutions to the scalar wave equation decay polynomially.

Key to Aretakis's instability proof is a  series of conservation laws for the wave equation along the event horizon of   general axisymmetric extremal black holes, and in particular extremal Kerr. The Aretakis instability has been further generalised by  Reall and  Lucietti \cite{lucietti_gravitational_2012} to non-axisymmetric perturbations and the  Teukolsky equation. This leads to  the remarkable conclusion that all extremal black holes are unstable to gravitational perturbations along their event horizon.
The  long-time ramifications of this in the full non-linear theory of general relativity is an  open problem.
 
\begin{figure}[ht]
        \centering
        \includegraphics[scale=0.4]{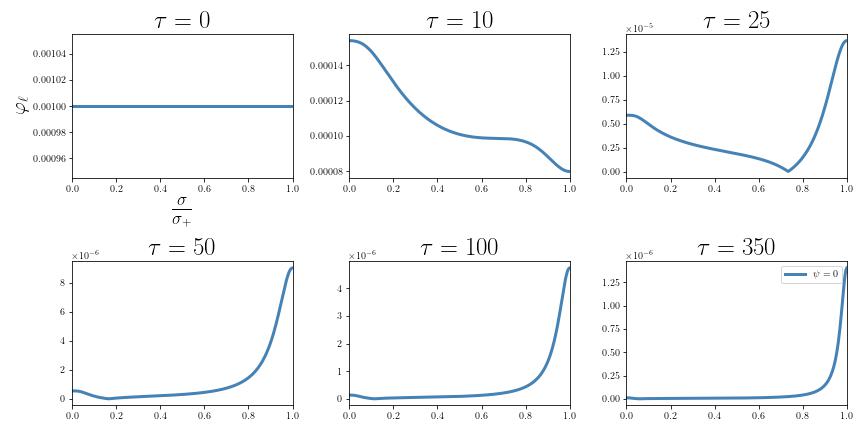}
        \caption{Initial data and subsequent snapshots of the $(\ell,m,s)=(0,0,0)$ perturbation  profile $\varphi_\ell(\tau,\sigma)$ for the extremal Kerr case $|a|=M$. The sharp gradients forming near the horizon on the right boundary are manifestations of the Aretakis instability.}
        \label{fig:initial_condition_aretakis}
\end{figure}

\begin{figure}[ht]
        \centering
        \includegraphics[scale=0.25]{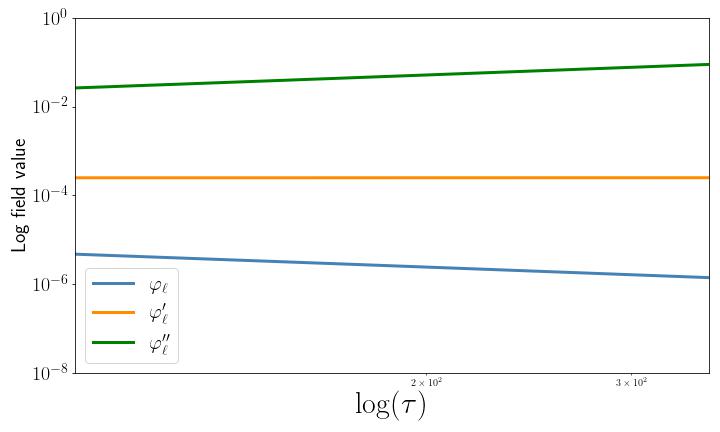}
        \includegraphics[scale=0.25]{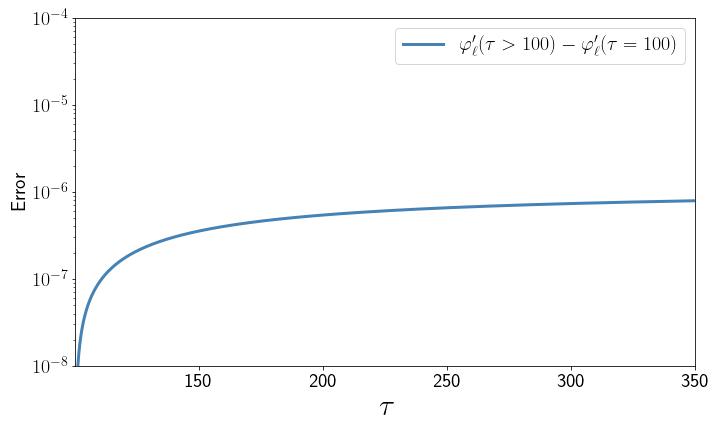}
        \caption{Decay for solutions of the wave equation on extremal Kerr spacetimes  $|a|=M$, for the fundamental axisymmetric mode $(l,m,s)=(0,0,0)$, exhibiting the Aretakis instability. The left plot shows that the field on the horizon decreases with time as $ \varphi_\ell(\tau,\sigma_+) \sim \tau^{-1}$, the first spatial derivative of the field is constant in time, $ \varphi'_\ell= \partial_\sigma \varphi_\ell(\tau,\sigma)|_{\sigma=\sigma_+} \sim \tau^0$, and the second spatial derivative of  the field increases with time as $\varphi''_\ell= \partial_\sigma^2 \varphi_0(\tau,\sigma)|_{\sigma=\sigma_+} \sim \tau$. The results were calculated using a pseudo-spectral grid of 96 points. The right plot shows the error in the conserved first derivative of the field.}
        \label{fig:Aretakis_results}
\end{figure}

In this section, we have  confirmed that our approach conserves the Aretakis charge on the horizon and can correctly reproduce the related instability \cite{aretakisHorizonInstabilityExtremal2013,aretakisDynamicsExtremalBlack2018}
for axisymmetric and non-axisymmetric perturbations of extremal Kerr spacetime. 
Our results in Fig.~\ref{fig:Aretakis_results} show that the $\ell=0$ mode of the field decays as
$\varphi_\ell(\tau,\sigma_+) \sim \tau^{-1}$ with time $\tau$, the first spatial derivative  $ \partial_\sigma \varphi_\ell(\tau,\sigma)|_{\sigma=\sigma_+} $    (which amounts to the conserved Aretakis charge) is constant in time, and the second spatial derivative  grows  as $ \partial_\sigma^2 \varphi_\ell(\tau,\sigma)|_{\sigma=\sigma_+} \sim \tau$, in accordance with Aretakis' theorem.

To compute the results in Fig.~\ref{fig:Aretakis_results} we used a similar set up to that used for computing the late-time tails outlined in the previous section. The main differences between the two setups is that we are computing the Aretakis instability for extremal Kerr, so the  spin parameter is set to $\chi = \frac{1}{2}$, that is $|a| = M$. Additionally,  the initial data of the field used for the Aretakis instability was a constant function, as shown in Fig.~\ref{fig:initial_condition_aretakis}. Aside from these differences, the setup used for the Aretakis instability is the same as for Price tails.

\begin{figure}[ht]
        \centering
        \includegraphics[scale=0.25]{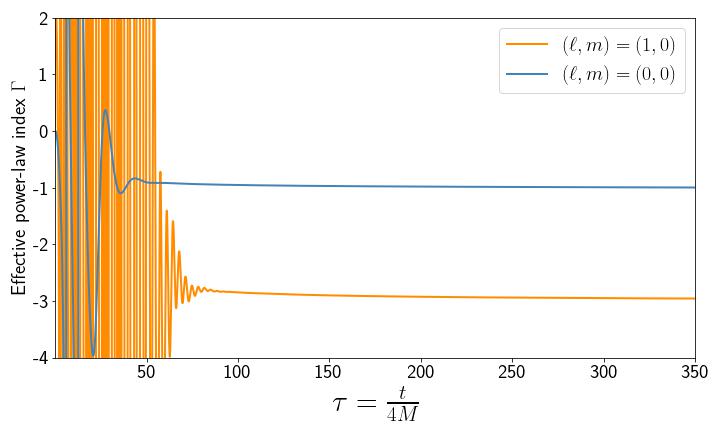}
        \includegraphics[scale=0.25]{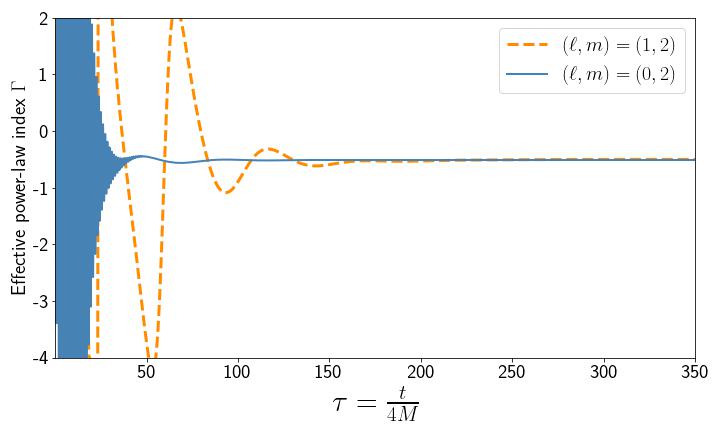}
        \caption{Late-time tails for the extremal Kerr case for spin $s=0$ fields. The left plot for axisymmetric mode $m=0$  shows that the field decays as  $ \varphi_\ell(\tau,\sigma_+) \sim \tau^{\ell-2l-1}$ on the horizon, in accordance with Aretakis' theorem. The right plot  for non-axisymmetric mode $m=2$ shows that the field decays as $ \varphi_\ell(\tau,\sigma_+) \sim \tau^{-1/2}$, in accordance with predictions made  by  Casals, Gralla and Zimmerman \cite{grallaTransientInstabilityRapidly2016,casalsHorizonInstabilityExtremal2016,grallaCriticalExponentsExtremal2018}.
}
        \label{fig:Aretakis_results_tails}
\end{figure}

\begin{figure}[ht]
        \centering
        \includegraphics[scale=0.25]{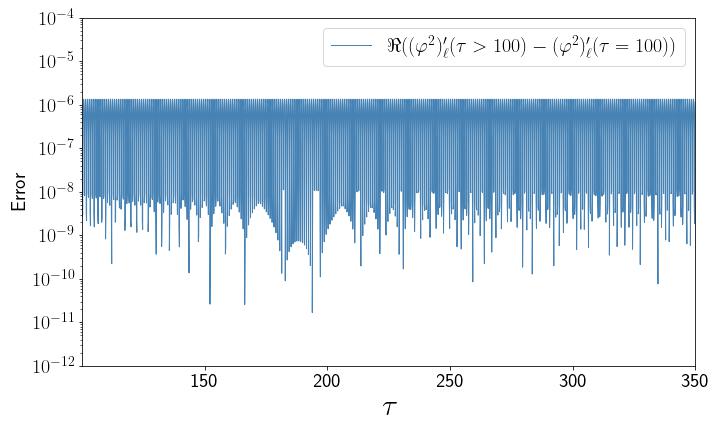}
        \includegraphics[scale=0.25]{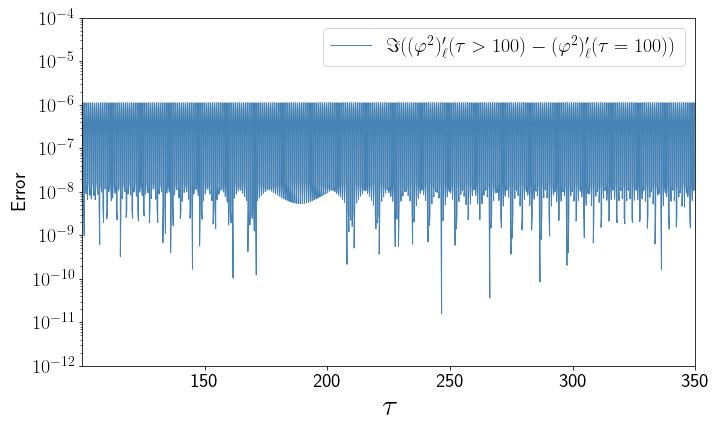}
        \caption{Consevation of a non-axisymmetric Aretakis charge for solutions of the wave equation on extremal Kerr spacetimes  $|a|=M$. The plot shows the error in the real and imaginary part of the charge  $ (\varphi^2_\ell)' = \partial_\sigma \varphi^2_\ell(\tau,\sigma)|_{\sigma=\sigma_+} \sim \tau^0$  for  a non-axisymmetric mode $(l,m,s)=(2,2,0)$, which is conserved on the future horizon $\mathscr{H}^+$ as predicted by Gralla et al. \cite{grallaTransientInstabilityRapidly2016,casalsHorizonInstabilityExtremal2016,grallaCriticalExponentsExtremal2018}.
The results were calculated using a pseudo-spectral grid of 96 points.}
        \label{fig:Aretakis_nonaxis_energy}
\end{figure}

For the extremal Kerr case, Fig.~\ref{fig:Aretakis_results_tails} shows that late-time tails for axisymmetric modes $(l,m,s)$ $=(0,0,0)$ and $(0,1,0)$   decay as  $ \varphi_\ell(\tau,\sigma_+) \sim \tau^{-2\ell-1}$ and that the Aretakis charge e  $ (\varphi_\ell)' = \partial_\sigma \varphi_\ell(\tau,\sigma)|_{\sigma=\sigma_+} $  is conserved on the future horizon $\mathcal{H^+}$, in accordance with Aretakis' theorem. In the non-axisymmetric case, superradiance leads to an enhanced growth rate of the Aretakis instability. Fig.~\ref{fig:Aretakis_results_tails}   shows that tails for non-axisymmetric modes $(l,m,s)=$ $(0,2,0)$ and $(1,2,0)$ decay as $ \varphi_\ell(\tau,\sigma_+) \sim \tau^{-1/2}$ on the horizon.
Fig.~\ref{fig:Aretakis_nonaxis_energy} shows that the Aretakis charge   $ (\varphi^2_\ell)' = \partial_\sigma \varphi^2_\ell(\tau,\sigma)|_{\sigma=\sigma_+} $  for a non-axisymmetric mode $(l,m,s)=$ $(2,2,0)$ is conserved  on $\mathcal{H^+}$. This is in accordance with  predictions  made by Casals, Gralla and Zimmerman \cite{grallaTransientInstabilityRapidly2016,casalsHorizonInstabilityExtremal2016,grallaCriticalExponentsExtremal2018} using Laplace transform techniques.
   
\newpage


\newpage
\section*{Acknowledgments}
We thank Yannis Angelopoulos for suggesting studying the Aretakis instability for non-axisymmetric modes. We also thank Sam Dolan for suggesting the use of Clebsch-Gordan coefficents to reduce the Teukolsky equation to  1+1 form and  Leor Barack for providing expressions of this equation in null coordinates. Finally, we thank Rodrigo Panosso Macedo and Juan A. Valiente-Kroon for useful discussions on  hyperboloidal compactification and the minimal gauge.
\newpage

\bibliography{Paper}



        
        
        

\end{document}